\documentclass[12pt,preprint2]{aastex}

\usepackage{overpic}
\usepackage{multirow}
\slugcomment{To appear in the Astrophysical Journal}
\shorttitle{BLAST}
\shortauthors{The Blast collaboration}

\newcommand{\degree}{\ensuremath{^\circ}}

\newcommand{\blast}{{BLAST}}
\newcommand{\blastfs}{{BLAST03}}
\newcommand{\blastkir}{{BLAST05}}
\newcommand{\blastmcm}{{BLAST06}}
\newcommand{\qvec}[2]    {\left(\begin{array}{c}
                                #1 \\
                                #2
                        \end{array}\right)}

\begin{document}

\title{The Balloon-borne Large Aperture Submillimeter Telescope: \blast}

\author{E.~Pascale,\altaffilmark{1,2}
	P.~A.~R.~Ade,\altaffilmark{2}
	J.~J.~Bock,\altaffilmark{3,4}
        E.~L.~Chapin,\altaffilmark{5}
        J.~Chung,\altaffilmark{1,5}
	M.~J.~Devlin,\altaffilmark{6}
	S~Dicker,\altaffilmark{6}
	M.~Griffin,\altaffilmark{2}
	J.~O.~Gundersen,\altaffilmark{7}
        M.~Halpern,\altaffilmark{5}
        P.~C.~Hargrave,\altaffilmark{2}
	D.~H.~Hughes,\altaffilmark{8}
	J.~Klein,\altaffilmark{6}
	C.~J.~MacTavish,\altaffilmark{1}
	G.~Marsden,\altaffilmark{5}
        P.~G.~Martin,\altaffilmark{9,10}
	T.~G.~Martin,\altaffilmark{1}
	P.~Mauskopf,\altaffilmark{2}
	C.~B.~Netterfield,\altaffilmark{1,10}
        L.~Olmi,\altaffilmark{11,12}
	G.~Patanchon,\altaffilmark{5,13}
	M.~Rex,\altaffilmark{6}
        D.~Scott,\altaffilmark{5}
	C.~Semisch,\altaffilmark{6}
	N.~Thomas,\altaffilmark{7}
	M.~D.~P.~Truch,\altaffilmark{14}
	C.~Tucker,\altaffilmark{2}
        G.~S.~Tucker,\altaffilmark{14}
	M.~P.~Viero,\altaffilmark{10}
	D.~V.~Wiebe\altaffilmark{1}}

\altaffiltext{1}{Department of Physics, University of Toronto, 60 St.~George Street, Toronto, ON M5S~1A7, Canada; {\url{enzo@physics.utoronto.ca}}}

\altaffiltext{2}{Department of Physics \& Astronomy, Cardiff University, 5 The Parade, Cardiff, CF24~3AA, UK}

\altaffiltext{3}{Jet Propulsion Laboratory, Pasadena, CA 91109-8099}

\altaffiltext{4}{Observational Cosmology, MS 59-33, California Institute of Technology, Pasadena, CA 91125}

\altaffiltext{5}{Department of Physics \& Astronomy, University of British Columbia,
		 6224 Agricultural Road, Vancouver, BC V6T~1Z1, Canada}

\altaffiltext{6}{Department of Physics and Astronomy, University of Pennsylvania, 209 South 33rd Street, Philadelphia, PA 19104}

\altaffiltext{7}{Department of Physics, University of Miami, 1320 Campo Sano Drive, Coral Gables, FL 33146}

\altaffiltext{8}{Instituto Nacional de Astrof\'isica \'Optica y Electr\'onica (INAOE), Aptdo. Postal 51 y 72000 Puebla, Mexico}

\altaffiltext{9}{Canadian Institute for Theoretical Astrophysics, University of Toronto, 60 St.~George Street, Toronto, ON M5S~3H8, Canada}

\altaffiltext{10}{Department of Astronomy \& Astrophysics, University of Toronto, 50 St.~George Street, Toronto, ON M5S~3H4, Canada}

\altaffiltext{11}{Istituto di Radioastronomia, Largo E. Fermi 5, I-50125, Firenze, Italy}

\altaffiltext{12}{University of Puerto Rico, Rio Piedras Campus, Physics Dept., Box 23343, UPR station, San Juan, Puerto Rico}

\altaffiltext{13}{Laboratoire APC, 10, rue Alice Domon et L{\'e}onie Duquet 75205 Paris, France}

\altaffiltext{14}{Department of Physics, Brown University, 182 Hope Street, Providence, RI 02912}

\begin{abstract}
The Balloon-borne Large Aperture Submillimeter Telescope (\blast) is a
sub-orbital surveying experiment designed to study the evolutionary
history and processes of star formation in local galaxies (including
the Milky Way) and galaxies at cosmological distances. The BLAST
continuum camera, which consists of 270 detectors distributed between
3 arrays, observes simultaneously in broad-band (30\%)
spectral-windows at 250, 350, and 500\,\micron. The optical design is
based on a 2\,m diameter telescope, providing a diffraction-limited
resolution of 30\arcsec\ at 250\,\micron. The gondola pointing system
enables raster mapping of arbitrary geometry, with a repeatable
positional accuracy of ${\sim}$\,30\arcsec; post-flight pointing
reconstruction to $\lesssim 5$\arcsec\ rms is achieved.  The on-board
telescope control software permits autonomous execution of a
pre-selected set of maps, with the option of manual override.  In this
paper we describe the primary characteristics and measured in-flight
performance of \blast\@. \blast\ performed a test-flight in 2003 and
has since made two scientifically productive long-duration balloon
flights: a 100-hour flight from ESRANGE (Kiruna), Sweden to Victoria
Island, northern Canada in June 2005; and a 250-hour,
circumpolar-flight from McMurdo Station, Antarctica, in December 2006.
\end{abstract}


%%%%%%%%%%%%%%%%%%%%%%%%%%%%%%%%%%%%%%%%%%%%
\section{Introduction}     \label{sec:intro}
%%%%%%%%%%%%%%%%%%%%%%%%%%%%%%%%%%%%%%%%%%%%

We have built and flown a balloon-borne submillimeter observatory,
designed to study the star formation in our Galaxy, in nearby resolved
galaxies, and in high-redshift starburst galaxies by observing at
several hundred microns, a wavelength range which is not available
from the ground.

The Balloon-borne Large Aperture Submillimeter Telescope (\blast) is a
stratospheric balloon-borne 2\,m telescope which observes the sky with
bolometric detectors operating in three 30\% wide bands at 250, 350,
and 500\,\micron.  The diffraction-limited optics are designed to
provide \blast\ with a resolution of $30\arcsec$, $42\arcsec$, and
$60\arcsec$ at the three wavebands, respectively.  The detectors and
cold optics are adapted from those to be used on the SPIRE instrument
on {\sl Herschel} \citep{grif03}.

\blast\ addresses important Galactic and cosmological questions
regarding the formation and evolution of stars, galaxies and clusters
\citep{devlin04} by providing the first large-area
(${\sim}$\,0.8--200\,deg$^2$) surveys of unique spectral-coverage,
angular resolution and sensitivity.

The primary scientific goals of \blast\ are: (i) to conduct
confusion-limited and shallower wide-area extragalactic surveys to
constrain the redshift-distribution, star formation history, and
evolution of optically-obscured luminous galaxies by measuring
photometric-redshifts \citep[derived from the \blast\ colors and
complementary data,][]{hughes02}; (ii) to study the spatial clustering
of this population; (iii) to improve our understanding of the earliest
stages of star formation by determining the physical properties and
mass-function of cold pre-stellar cores and the efficiency of star
formation within different Galactic environments; and (iv) to
investigate the nature and structure of the interstellar medium by
making the highest resolution maps to date of diffuse Galactic
emission at these wavelengths.

The \blast\ bands bracket the peak of the thermal radiation emitted by
dust at temperatures of around 10\,K\@.  Assuming that the temperature
range of dust in submillimeter galaxies is ${\sim}$\,30--60\,K,
\blast\ can explore redshifts from 1 to 5 \citep{hughes02}.  The
primary advantage of \blast\ over existing submillimeter bolometer
arrays such as {SCUBA} and {SHARC} is its enhanced sensitivity at
wavelengths $\le 500$\,\micron\ (where the far-IR background peaks) due
to the increased atmospheric transmission at balloon altitudes. With
sensitivities of 250\,mJy\,s$^{1/2}$, the \blast\ mapping speed is 10
times faster than the design goal for {SCUBA}\,2 at 500\,\micron\ 
\citep{holland06} and more than 100 times the mapping speed of the
pioneering flights of the FIR balloon-borne telescope {PRONAOS}
\citep{lam98}.  \blast\ is complemented at shorter wavelengths by
surveys from {\it IRAS}, {\it ISO}, {\it Spitzer} and {\it Akari}, and
at longer wavelengths by SCUBA~2, LMT, and ALMA, to constrain spectral
energy distributions\@. {\it Spitzer\/}'s higher resolution will also
provide accurate astrometry for many of the \blast\ sources.  \blast\
complements the {\it Herschel}\ satellite by testing detectors and
filters which are similar to those to be used in the SPIRE instrument
\citep{grif03}. Furthermore, the results from \blast\ will be
available early enough to influence the design (depth and area) of
future {SPIRE} surveys.  In addition, \blast\ will complement the
large-area spectroscopic Galactic surveys of {\it SWAS}, provide
submillimeter targets for the Fabry-Perot spectrograph {SAFIRE} on
{\it SOFIA}, and impact the design of the scientific case for the next
generation of submillimeter surveys from space (e.g.\ {\it SPICA} and
{\it SAFIR}, and later {\it SPIRIT} and {\it SPECS}).

\blast\ has made three flights to date. A 24-hour test flight
(\blastfs) from Fort Sumner, NM, in September 2003 demonstrated the
performance of the instrument sub-systems.

\blast\ made its first science flight (\blastkir) in June of 2005,
flying on a $1.1 \times 10^6$\,m$^3$ balloon from the Swedish Space
Corporation base of ESRANGE, near Kiruna, Sweden to Victoria Island in
northern Canada.  During this 4.5-day flight at an average altitude of
38\,km, the instrument performed well, except for degraded optical
performance. This was possibly due to a failure of structural elements
in the carbon fiber mirror. The result was decreased resolution, which
had a significant impact on our ability to achieve our extragalactic
science goals\@.  \blast\ acquired 100 hours of data on Galactic
targets, providing some of the first arcminute resolution images at
these wavelengths \citep{Chapin2008}.  In this flight, surveys of five
star forming regions were conducted, including the well-studied
Cygnus-X field, three regions of intermediate/high-velocity cirrus,
the Cassiopeia-A supernova remnant, and several individual bright
targets.

\blast\ made its second science flight (\blastmcm) in December, 2006,
flying from the Williams Field Long Duration Balloon (LDB) facility
near McMurdo Station in Antarctica.  In this flight, the instrument
met its performance goals in resolution, sensitivity, and pointing.
The flight was terminated after 250 hours of data were acquired.
Unfortunately, an anomaly with the parachute separation system
prevented the parachute from being released from the gondola after
landing.  The parachute dragged \blast\ for 200\,km before it came to
rest in a crevasse field on the Antarctic Plateau.  While most of the
instrument was destroyed, the hard drives containing the data were
found nearby, and the complete dataset recovered.  In this flight,
\blast\ conducted shallow (10\,deg$^2$) and deep, confusion-limited
(0.8\,deg$^2$) extragalactic surveys of the Chandra Deep Field South
(CDF-S) as well as a 10\,deg$^2$ region near the South Ecliptic Pole.
A Galactic survey in Vela mapped 52\,deg$^2$ and 200\,deg$^2$ of sky
in deep and shallow observations, respectively.  Surveys of several
Galactic and extragalactic selected targets were also conducted.

The telescope and optics are described in Section~\ref{sec:telescope},
while Sections~\ref{sec:rec}
and~\ref{sec:cryo} focus on detector arrays and
cryogenics.  Sections~\ref{sec:gondola} through \ref{sec:psens} provide detailed information regarding the
gondola, command and control, and pointing systems.  Sections \ref{sec:thermal} through \ref{sec:performance} provide
overviews of the unique thermal and power requirements, and a
description of the pointing performance.

%%%%%%%%%%%%%%%%%%%%%%%%%%%%%%%%%%%%%%%%%%%%%%%%%%%
\section{Telescope} \label{sec:telescope}
%%%%%%%%%%%%%%%%%%%%%%%%%%%%%%%%%%%%%%%%%%%%%%%%%%%

\begin{figure*}
        \centering
	\includegraphics[width=0.8\linewidth]{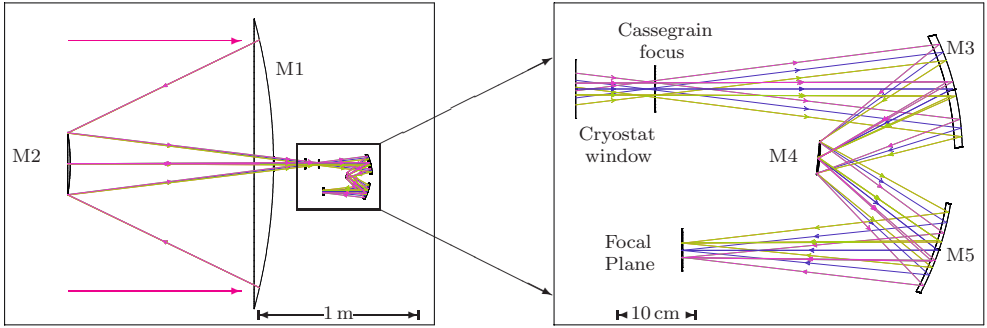}
\caption{The optical layout of the \blastkir\ telescope and receiver
is shown on the left and the 1.5\,K optics, located within the
cryostat, are shown in expanded view on the right.  The image of the
sky formed at the input aperture is re-imaged onto the bolometer
detector array at the focal plane.  The mirror M4 serves as a Lyot
stop defining the illumination of the primary mirror for each element
of the bolometer array.  The three wavelength bands are separated by a
pair of dichroic beamsplitters (not shown) which are arranged in a
direction perpendicular to the plane, between M5 and the focal plane.}
\label{fig:zeemax_design}
\end{figure*}

The basic \blast\ optical configuration is shown in
Figure~\ref{fig:zeemax_design}\@.  Incoming radiation is collected by
the telescope (M1 and M2 in the figure) which is located on the inner
frame of the gondola (described in Section~\ref{sec:gondola}), along
with the cryogenic receiver (Section \ref{sec:cryo})\@. The telescope
focus is located about 20\,cm behind the optical surface of the
primary.  Cold (1.5\,K) re-imaging optics (M3, M4, and M5) in the
receiver relay the sky image to the detector focal planes. A baffled
Lyot stop (M4) is located at the position of an image of the primary
mirror.  This blocks stray radiation due to scattering and
diffraction.  After M5 the radiation is divided into three spectral
bands through the use of two low-pass edge dichroic filters
\citep{ade06}.  An aperture at the center of the Lyot stop is made to
match the one at the center of the primary mirror to reduce loading.
This same hole accommodates a commandable thermal source, which
provides a stable and repeatable optical signal (calibration lamp) to
monitor detector responsivity drifts \citep{hargrave06}.  The optical
design is optimized to deliver a diffraction-limited image of the sky
at the detector focal planes, with 60 and 30\arcsec\ beams at
500 and 250\,\micron, respectively. The telescope designs of
\blastkir\ and \blastmcm\ are discussed in this section and the
relevant parameters can be found in Table~\ref{table:blast_optics}.

\subsection{Focal Plane}

The illumination of the Lyot stop and the main reflector depends
primarily on the properties of the feed-horns in the detector array.
\blast\ uses $2f\lambda$ spaced, smooth-walled conical feeds which are
similar to the SPIRE feeds. Their design and measured characteristics
have been described by~\citet{chatt03}, \citet{rownd03}, and
\citet{griffin02}.  The optical system design is based upon the
feed-horn nominal $f/5$ focal ratio and requires the chief-rays to be
perpendicular to the aperture of the horn, with all the beams
overlapping at the Lyot stop.  Knowledge of the fundamental propagated
electromagnetic mode (TE$_{11}$) is used in a
Zemax\footnote{\url{http://www.zemax.com/}} Physical Optics module to verify
the final design.

\subsection{\blastkir\ Optical Design}	

For the 2005 flight, the design incorporated a 2\,m carbon fiber
spherical primary mirror with a mass of 32\,kg and a surface rms of
2.4\,\micron.  It was designed and built by Composite Optics
Incorporated\footnote{Composite Optics Incorporated (COI), 9617
Distribution Avenue, San Diego, CA 92121.} as a technology prototype
for the 3.5\,m {\em Herschel\/} telescope.  An aluminum correcting
secondary mirror, with a diameter of 50\,cm, was designed to give a
diffraction-limited performance over a $14'\times7'$ field of view
(FOV) at the telescope focus.  The estimated antenna efficiency is
$\ge 80\%$; losses come from a combination of the rms surface
roughness of the primary and the quality of the re-imaging optics. Two
of the re-imaging cold elements, M3 and M5, are ellipsoidal mirrors;
M2 and the other cold element, M4, compensate for most of the
spherical aberration of the primary, whereas M3 and M5 contribute to
achieve the correct focal ratio.  The perimeter of the primary mirror
was structurally weak, forcing M2 to be suspended by a carbon fiber
structure attached through the hole in the primary mirror.  This
geometry resulted in ${\sim}$\,12\% blockage of the beam.
A detailed discussion  about \blastkir\ optical design and optimization can be found in~\citet{olmi01, olmi02}. 
The 2\,m primary mirror was
under-illuminated by setting the entrance pupil diameter to 188\,cm
and the reflector was strongly tapered, resulting in a field taper of
about $-15$\,dB. A Strehl ratio $> 0.96$ was achieved over the whole focal
plane, and far field beams had a first sidelobe at $-10$\,dB. A detailed analysis of the relation  between Strehl ratio and aperture efficiency, which is more commonly quoted for filled-aperture antennas, can be found in \citet{olmi07}.

\subsection{\blastmcm\ Optical Design}
In the 2006 Antarctic flight, \blast\ flew a Ritchey-Chr\'etien
telescope, with a 1.8\,m diameter, aluminum primary mirror with a mass
of 114\,kg.  Originally, the mirror was intended to be used only for
the test flight in 2003.  However, after the destruction of the carbon
fiber mirror on landing of \blastkir, the aluminum mirror was
re-machined to improve the quality of the surface from 8\,\micron\ to
better than 4\,\micron\ rms\@.  This machining was performed by the
Precision Engineering Group at Lawrence Livermore National Laboratory.
A 40\,cm diameter aluminum secondary was suspended by four carbon
fiber struts\footnote{ Innovative Composite Engineering, 139
  E. Columbia River Way, Bingen, WA 98605.}  (with a zero linear
thermal expansion coefficient) attached to the perimeter of the
primary mirror. This reduced the obscuration to ${\sim}$\,7\% compared to
the \blastkir\ design.  Simulations of in--flight detector loading
have shown a reduction of $\gtrsim 40$\,\% for this solution,
relative  to the  \blastkir\ configuration. 
In an attempt to further reduce loading, angled reflecting baffles
were installed on the inner surface of each strut to deflect most of
the obscured beam to the relatively cold sky.
The cold re-imaging optics form an ideal Offner relay. In this
configuration, M3, M4, and M5 are all spherical and share a common
center of curvature; M3 and M5 are concave while M4 is convex. In order
to achieve a diffraction-limited beam of 60\arcsec\ at 500\,\micron\ and
30\arcsec\ at 250\,\micron\ with a smaller diameter primary compared to
\blastkir, M1 was more aggressively illuminated. This was made possible
by reducing the diameter of the Lyot stop by about 12\%, leading to a
field taper of $-7.5$\,dB on M1 and $-17$\,dB sidelobes in the far field
beams.

\subsection{\blastmcm\ Focussing System}

The relative distance between the primary and the secondary mirrors
has a tolerance of one wavelength before significant image degradation
is introduced.  Thermal modeling indicated that diurnal temperature
fluctuations at balloon altitudes of the aluminum M1 and M2 mirrors
could have been as large as 10\degree\/C. The required correction in
the relative distance between M1 and M2 was calculated to be
50\,\micron/\degree\/C.

To correct for the thermal motion, we implemented a motorized system
for in-flight refocusing of the secondary.  M2 was coupled to its
mounting structure via 17-7PH stainless steel leaf springs. Three
stepper motor actuators\footnote{Ultra Motion, 22355 Route 48, \#21,
Cutchogue, NY 11935.} (2.12\,mm\,rev$^{-1}$ lead screw) drive the
mirror back and forth to correct for the focus position, as well as
adding tip/tilt capability to initially align M2 to M1.  Accurate
positioning was achieved with differential optical encoders ($\simeq$1
count \micron$^{-1}$ resolution) factory mounted to the motors, and
three DC Linear Variable Differential Transformers\footnote{Macro
Sensors, 7300 US Route 130 North, Building 22, Pennsauken, NJ 08110,
0.6\,\micron\ or better \mbox{repeatability}.} (LVDT)\@.  This system
allowed for $\pm$5\,mm of motion about the nominal telescope focus,
which was sufficient for accurate positioning.  The flight computer
commands each motor individually by communicating with the stepper
motor controllers over a RS485 bus.

Pre-flight focusing of the system was hampered by the fact that water
vapor absorption makes ground-based submillimeter observations in the
far-field of the telescope (about 10\,km) impossible.  To mitigate
this, the secondary mirror was physically offset to 3 positions so
that the focal plane was reimaged at 50, 100, and 150\,m from the
telescope.  The focusing system was then used to fine-tune the focus
at each position.  These results were extrapolated to determine the
far-field focus for the flight.

In flight, the focus of the system was verified and had to be manually
adjusted by maximizing the response when scanning over bright point
sources: a correction of 300\,\micron\ to the position set at launch
was required.  Subsequently, the temperatures of the primary and
secondary were monitored and the secondary automatically repositioned
to account for thermal variations that would produce a focus
displacement of 100\,\micron\ or more. The primary and secondary
mirror diurnal temperature fluctuations were $\pm1.5^\circ$C for most
of the flight.

%%%%%%%%%%%%%%%%%%%%%%%%%%%%%%%%%%%%%%%%%%%%%%
\subsection{Array Bandpass Characteristics}
%%%%%%%%%%%%%%%%%%%%%%%%%%%%%%%%%%%%%%%%%%%%%%

\begin{figure*}
        \centering
	\includegraphics[angle=-90, width=\linewidth]{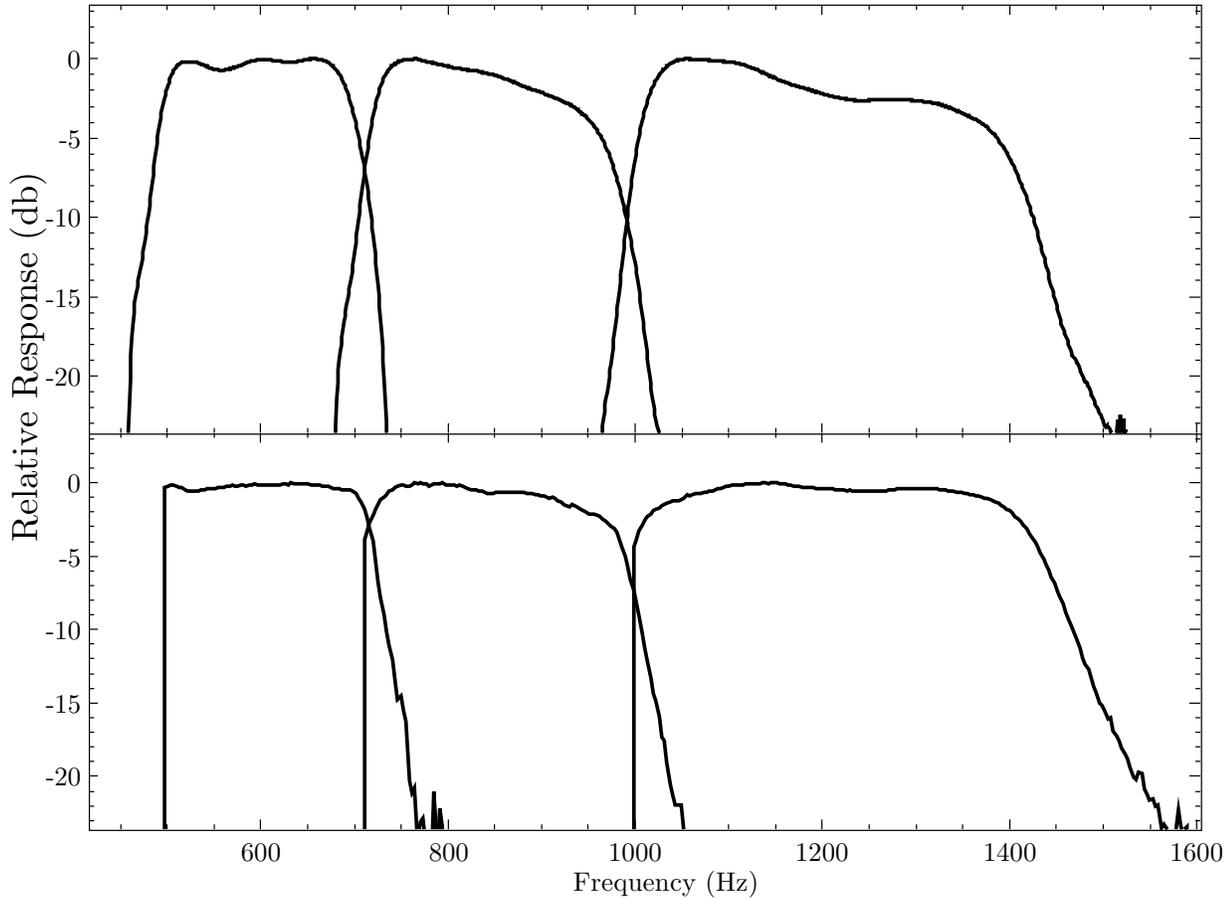}
	\caption{Relative spectral response of the three channels \blast\ channels. The \blastkir\ bands (top) have been measured with the filter stack in the cryostat. The \blastmcm\ (bottom)  bands are obtained by multiplying the profiles measured for each filter individually; the feed waveguide cut--off is assumed at the longer wavelength. A Fourier Transform Spectrometer has been used in both cases.}
\label{fig:band_pass}
\end{figure*}

Low-pass edge dichroic filters~\citep{ade06} split the incoming
radiation emerging from M5.  The first dichroic filter reflects
wavelengths shorter than 300\,\micron\ and transmits longer
wavelengths.  The bandpass for the 250\,\micron\ array is further
defined by a filter directly in front of the array, which reflects
wavelengths shorter than 215\,\micron, and by the waveguide cutoff at
the exit of each of the feedhorns.  For the 350\,\micron\ and
500\,\micron\ arrays, in each case, the band is defined at the short
wavelength end by the dichroic transmission and at the long
wavelength end by the waveguide cutoff. Each band has a 30\% width.
The filter stack frequency performance was evaluated with Fourier
Transform Spectroscopy. Bandpasses are shown in
Figure~\ref{fig:band_pass}.

%%%%%%%%%%%%%%%%%%%%%%%%%%%%%%%%%%%%%%%%%%%%%%%%%%%%%%%%%%%%%%%%
\section{Receiver} \label{sec:rec}
%%%%%%%%%%%%%%%%%%%%%%%%%%%%%%%%%%%%%%%%%%%%%%%%%%%%%%%%%%%%%%%%

The \blast\ focal plane consists of arrays of 149, 88, and 43
detectors at 250, 350, and 500\,\micron, respectively.  The arrays are
cooled to a temperature of $270$\,mK.  Each array element is a silicon
nitride micromesh ``spider web" bolometer \citep{bock96, bock98}.  The
detector arrays, the feeds, and the mounting scheme are based on the
SPIRE instrument design \citep{turner01}.  Each array hosts a number
of diagnostic channels: two dark bolometers (channels that have been
capped to avoid illumination), two thermistors, and one resistor.  The
detector parameters are summarized in Table~\ref{table:det_params}.

The detectors and the readout electronics are designed so that the
sensitivity is ultimately limited by the photon ``BLIP'' noise. To
achieve this performance, other contributions to the bolometer noise
have to be controlled. A bolometer is fundamentally limited by phonon
noise in the thermal link with the heat sink. The Noise Equivalent
Power (NEP) \citep{bock96} is NEP~$= \gamma\sqrt{4k_{\mbox{B}} T^2
G}$, where $G$ is the thermal conductance, $T$ is the bath temperature
and $\gamma$ takes into account the properties of the thermal link
between the detector and the bath. For a given background load $Q$,
the maximum sensitivity is achieved when
$G \sim Q/T$~\citep{mather_1984}.  The detectors have been
optimized for loads of 55, 40, and 30 pW at 250, 350, and
500\,\micron, respectively.

The detector transfer function is measured using cosmic ray hits, as described
in~\citet{crill03}. We have found typical time constants of $2$\,ms with a dispersion of 15\% across the arrays.

An additional advantage to the ``spider web'' bolometer
is the relatively small cross-section to cosmic rays.  Balloon flights
at high-latitudes have about 100 times the mid-latitude cosmic ray
flux. The detected cosmic ray flux was 1.5, 2, and 6 events per bolometer per minute during both science  flights at 
250, 350, and 500\,\micron, respectively.

\subsection{Readout Electronics}
	
\begin{figure*}
        \centering
	\includegraphics[angle=-90, width=\linewidth]{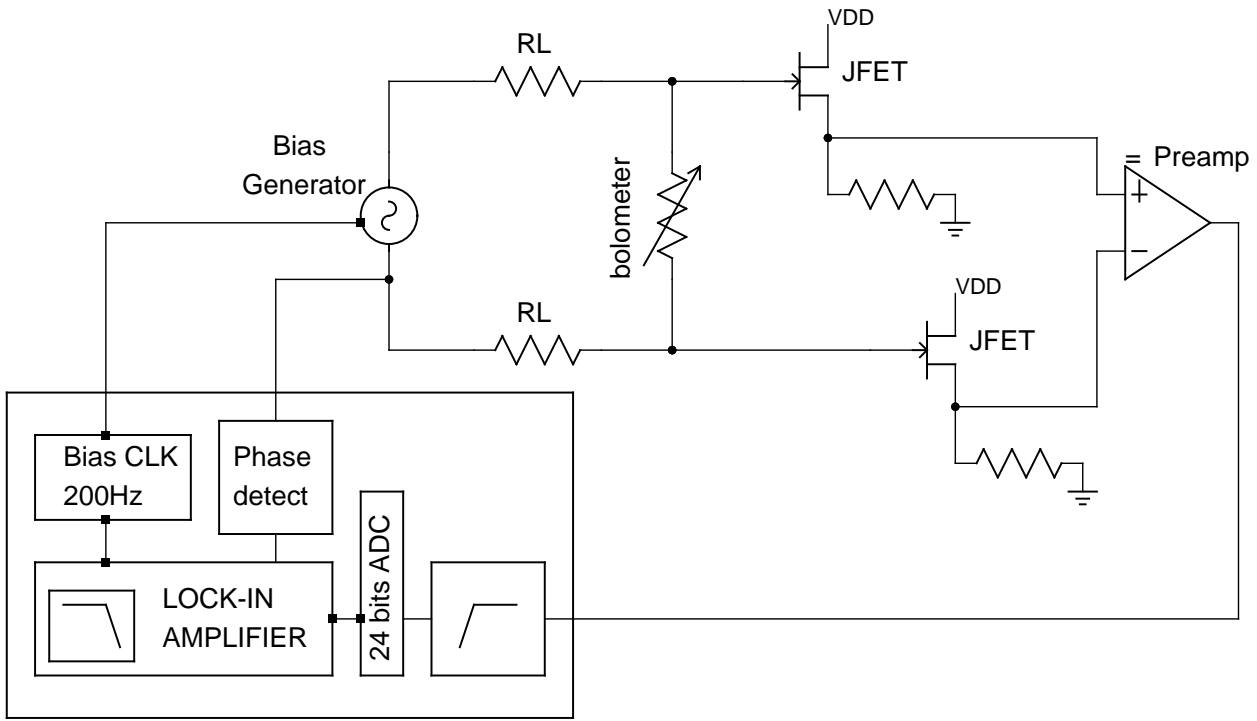}
	\caption{A schematic diagram of the detector readout
	electronics. A 200.32\,Hz sine wave biases each bolometer
	across two $7$\,M$\Omega$ load resistors (RL). JFET pairs
	buffer the detector signal to the warm electronics. The two
	signals are then differenced and digitized. A digital lock-in
	rectifies and filters the output sending it to the flight
	computers over the BLASTbus.}
	\label{fig:det_ro}
\end{figure*}

A schematic diagram of the readout electronics is shown in
Figure~\ref{fig:det_ro}.  Each bolometer is pre-amplified with a
Siliconix U401 differential JFET with 5--7\,nV\,Hz$^{-1/2}$ noise at
$\nu > 100\,\mbox{Hz}$. Compact 24 channel JFET modules are integrated
into the design of the cryostat, allowing us to sink their power
dissipation ($240\,\mu$W per pair) to the vapor-cooled shield (see
Section \ref{sec:cryo}) and decrease the load on the helium bath by
approximately 60\,mW. The JFETs are operated at a temperature of
${\sim}$\,145\,K.  Their output is then amplified using the Analog
AD624 instrumentation amplifier (5\,nV\,Hz$^{-1/2}$) with a 100\,Hz
wide band-pass filter, centered at 208\,Hz\@.
To mitigate the effects of $1/f$ noise, the detectors are AC biased with a sine
wave at 200.32\,Hz. A reference square wave at this frequency is
generated by dividing down the main 32\,MHz clock serving the readout
electronics, making the bias synchronous with the sampling. This
signal is then split into three paths (one per band), stabilized to a
voltage amplitude that is selectable from 0 to 300\,mV with a resolution
of 7 bits, and filtered into a sine wave using low-noise operational
amplifiers (Analog OP470). The bias is delivered to the detectors and
a reference is sent back to the Data Acquisition System (DAS) where
a digital lock-in is implemented on the DSP (Analog ADSP21062) in the
readout electronics. The bolometric AC signals and the bias reference
are digitized with a 24\,bit $\Sigma\Delta$ (Burr-Brown ADS1252) fed by
a low-noise dual instrumental amplifier (Burr-Brown INA2128) and
sampled at 10\,kHz. The reference is phase-locked and a numerical sine
wave is generated in-phase at the bias frequency (the clock is
shared). The rectified signal is then low-passed with a 4 stage
boxcar filter having a first null at 50.08\,Hz, and decimated to match the
DAS sampling frequency (100.16\,Hz).
 
The resulting flat-phase filter is well approximated by a Gaussian
and is computationally efficient in the time domain, allowing up to 25
independent channels to be locked-in on a single DSP.  Only the real
part of the signal is sampled, hence the relative phase between signal
and reference must be set manually. Since the phase difference is a
function of the bolometer impedance, it needs to be adjusted in flight
along with bias levels.  This is achieved by maximizing the signal
from the pulsed calibration lamp at the Lyot stop.

The overall data acquisition electronics noise is kept below the
estimated photon noise and  
provides  stability to low frequencies ($<30$\,mHz), which allows the
sky to be observed 
in a slowly-scanned mode. 

\subsection{Detector Noise}

The instrument is designed to be limited by photon noise.  The error
budget, which drove the telescope and readout electronics design,
accounted for (i) the photon noise arising from astrophysical
backgrounds and emissivity of the optical surfaces, (ii) the bolometer
noise budget, and (iii) the readout noise. A Noise Equivalent Flux
Density (NEFD) of 250\,mJy\,s$^{1/2}$ was expected from this analysis.

\begin{figure*}
        \centering
        \includegraphics[angle=-90, width=\linewidth]
        {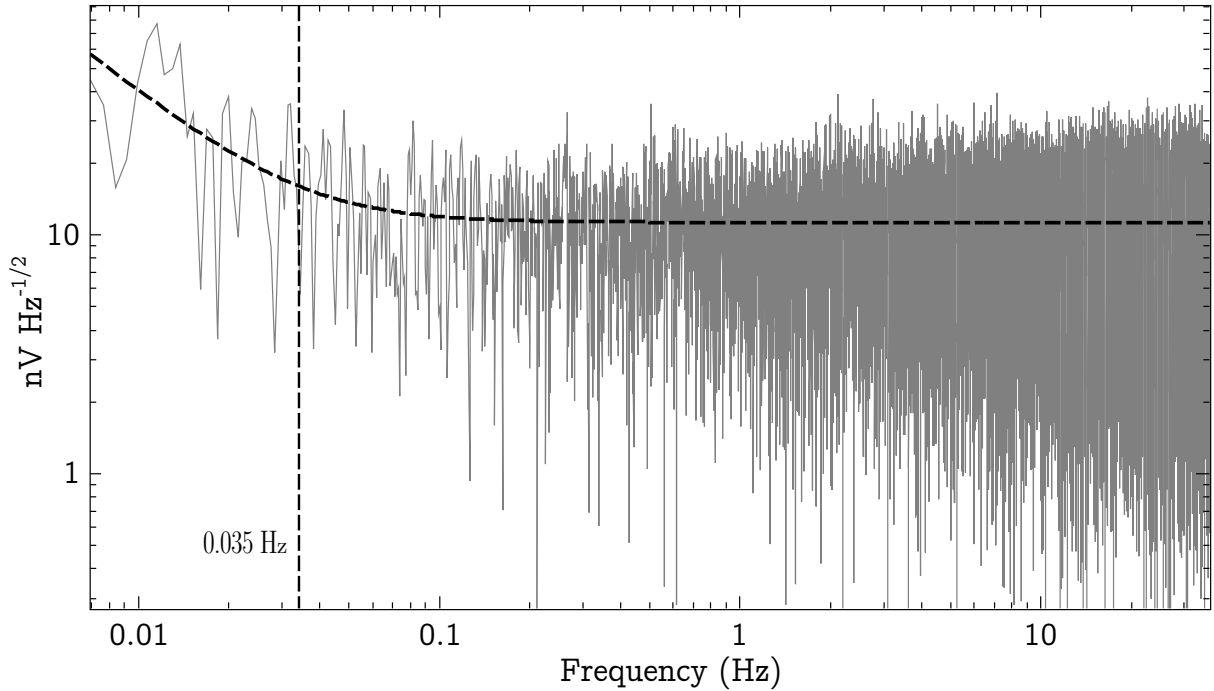} 

	\caption{A 250\,\micron\ detector noise power spectrum density (PSD).
The time stream was acquired during the \blastkir\ flight and analyzed as explained in \citet{patanchon2007}:  (i) the time stream has been deglitched, (ii) the detector and electronics transfer functions have been deconvolved from the data, and (iii) a common mode signal, synchronous to each  channel in a array, has been removed in the time domain.
The PSD is shown before any gain. The dashed line is a brown noise fit to the PSD which shows a white noise level at 11\,nV\,Hz$^{-1/2}$ and a 1/f ``knee'' at 35\,mHz (indicated by the vertical dashed line).}
\label{fig:noise}
\end{figure*}

The voltage noise is determined by taking the Power Spectral Density
(PSD) of the time stream acquired in flight and it is shown in
Figure~\ref{fig:noise} for a representative 250\,\micron\ detector
during the \blastkir\ flight. Before taking the PSD, each time stream
is processed as described in \citet{truch2007} and
\citet{patanchon2007}, which involves deglitching, deconvolution of
the bolometer and electronics transfer functions, and removal of
common mode signals, synchronous to each detector in a array.

The voltage noise is well described by a brown noise model. The white
noise level is 11\,nV\,Hz$^{-1/2}$ and the 1/f ``knee'' is at
${\sim}$\,35\,mHz. Using the responsivities in \citet{truch2007}, we
obtain a noise of ${\sim}$\,15\,MJy\,sr$^{-1}$\,Hz$^{-1/2}$. A similar
analysis at the longer wavelengths gives ${\sim}$\,7 and
4\,MJy\,sr$^{-1}$\,Hz$^{-1/2}$ at $350$ and 500\,\micron,
respectively.  Within a single array the pixel noise varies by
20\%. 

Measured noise levels are consistent with the design target for the NEFD
for nominal beam sizes.

%%%%%%%%%%%%%%%%%%%%%%%%%%%%%%%%%%%%%%%%%%%%%%%%%%%%%%%%%%%%%%%%%%%%%%%%%%%%
\section{Cryogenics} \label{sec:cryo}
%%%%%%%%%%%%%%%%%%%%%%%%%%%%%%%%%%%%%%%%%%%%%%%%%%%%%%%%%%%%%%%%%%%%%%%%%%%%
%CAN: overview

\begin{figure*}
        \centering
	\includegraphics[width=0.9\linewidth]{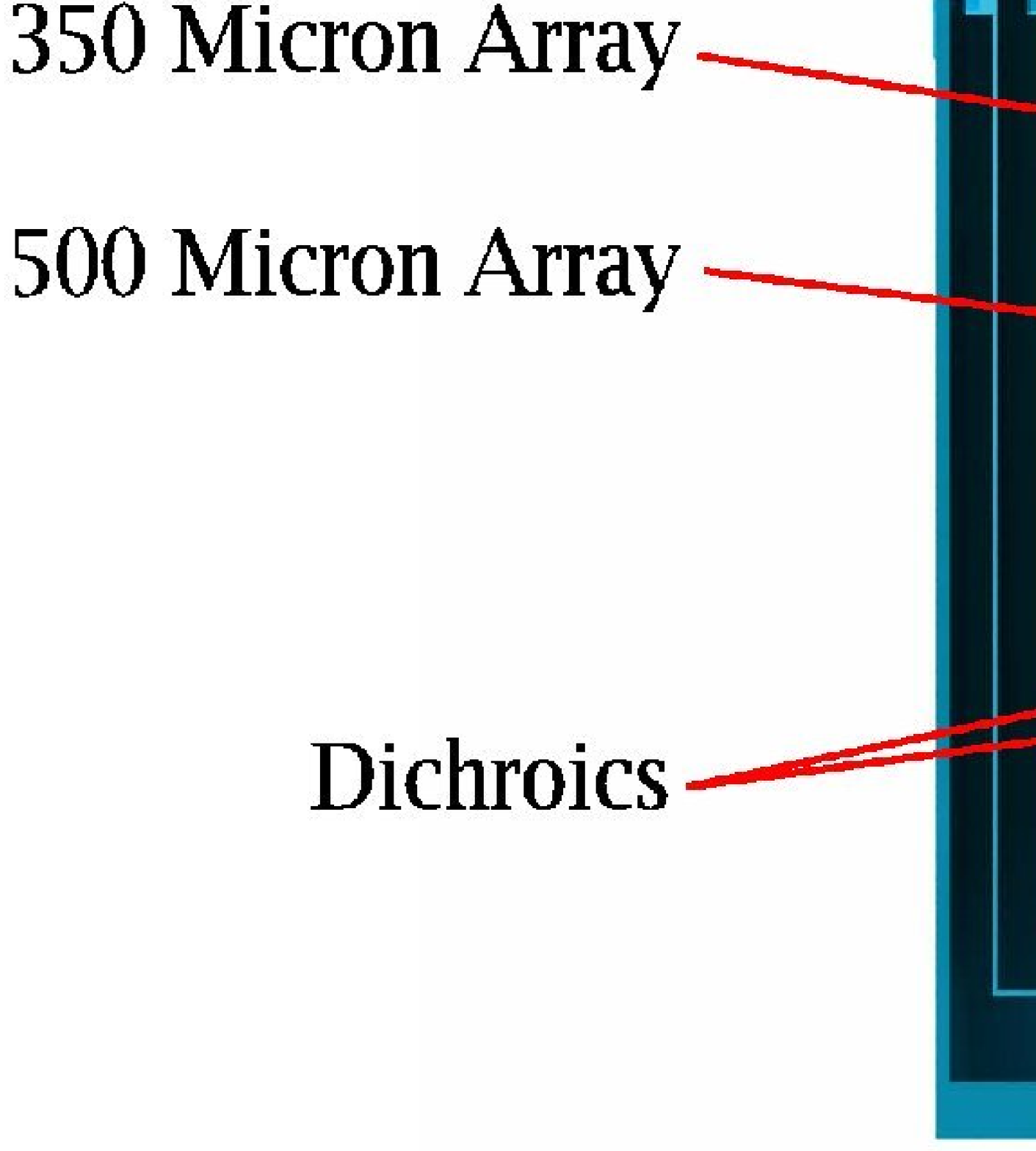}
	\caption{Cutaway model of the \blast\ receiver showing the
	optics box.  The $^3$He refrigerator is omitted for clarity.
	The cryostat is held to the telescope structure via bolts in 
	the top flange and jack screws around the perimeter near the base.}
	\label{cryostat}
\end{figure*}

The cryostat houses most of the receiver system (Figure
\ref{cryostat}) and was fabricated by Precision
Cryogenics.\footnote{\url{http://www.precisioncryo.com/}}  It is constructed
of aluminum and G10 woven fiberglass reinforced resin epoxy and is
maintained at vacuum to prevent thermal convection.  Each thermal
stage is light-tight except the optical path, which is thermally
protected with infrared blocking filters \citep{truch-thesis}.

Liquid baths of nitrogen and helium maintain the temperatures of the
77\,K and 4.2\,K stages.  In flight, an absolute pressure
regulator\footnote{Tavco, Inc, 20500 Prairie St., Chatsworth, CA.} is
used to maintain approximately 1\,atm above each bath.  A vapor-cooled
shield (VCS) is located between the nitrogen and helium stages which
is cooled by boil-off gas from the helium bath and reduces the thermal
loading on the helium stage.  Radiative and conductive loading is
7.2\,W on the nitrogen tank and 71\,mW on the helium tank.  An
additional effective load of 9\,mW and 7\,mW on the helium tank is due
to the liquid drawn into the pumped-pot by the capillary (discussed
below), and heating required to cycle the $^3$He refrigerator,
respectively.  The cryostat holds 43\,L of nitrogen and 32\,L of
helium and has an 11.5 day hold time.

A liquid He pumped-pot maintains the optics box
at 1.5\,K.  A long, thin capillary connects the helium bath to the
pumped-pot with a volume of ${\sim}$\,100\,mL.  A pump line between the
pot and the exterior of the cryostat keeps the pot at near vacuum,
maintaining the temperature at 1.5\,K.  The pressure difference between
the helium bath and the pot forces helium into the pot.
On the ground, a vacuum pump is used to evacuate the pump line; at
float altitude, the pump line is open to the atmosphere, which is at
a pressure $< 0.01$\,atm\@.

The bolometers and feed horns are maintained at $< 300$\,mK by the
closed-cycle $^3$He refrigerator. The $^3$He is condensed by the
1.5\,K pumped-pot and collects in the $^3$He cold stage.
The $^3$He refrigerator is
able to extract 5\,J of energy and needs to be recycled every 48--60 hours.
The recycling process takes less than 2.5 hours.

The cryostat performed as expected during both science flights.  The
nitrogen and helium baths maintained constant temperature and the VCS
fluctuated by less than 3\,K due to the $^3$He refrigerator cycles.
The pumped-pot maintained a temperature below 1.8\,K with fluctuations
less than 10\,mK\@.  During $^3$He refrigerator cycles the pumped-pot
temperature rose as high as 2.5\,K\@.  The bolometer temperature was
maintained below 300\,mK with less than 1\,mK fluctuations on hour
time scales.  The first science flight was too short to test hold
time.  In the second 12.5 day flight, the helium tank was exhausted
11.5 days after reaching float altitude.  The nitrogen tank was not
exhausted before termination.
\begin{figure*}
        \centering
	\includegraphics[width=\linewidth]{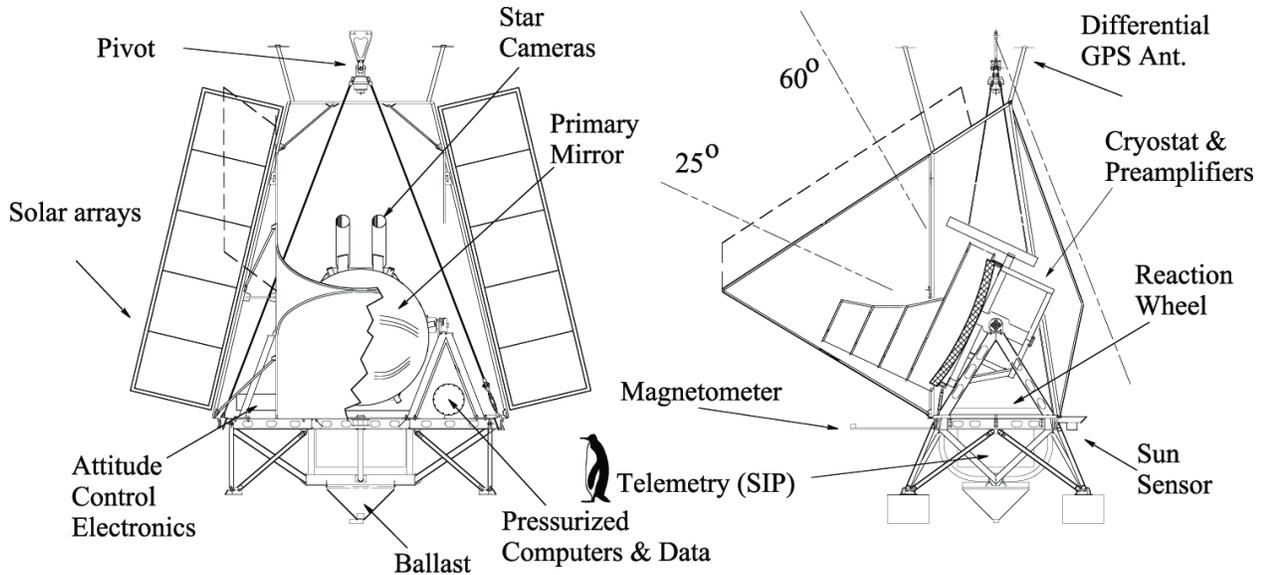}
	\caption{Front and side schematic drawings of the BLAST gondola.
  A 1\,m tall Emperor penguin is shown for scale. The height of the pivot is set by
 the size of the launch vehicle. The width of the gondola {\it
 without\/} deploying the solar panel arrays matches the width of the
 laboratory doors. The inner frame, which can be pointed in elevation,
 consists of the two star cameras, the telescope and its light baffle,
 the receiver cryostat and associated electronics.  The telescope baffle
 shown at right, which was used in 2005, was replaced by a smaller
 system in 2006.  The CSBF solar panels and transmitting antennae are
 suspended below the structure shown here. The lines marked 25\degree\
 and 60\degree\ show the useful range of orientation of the optic axis.
 The dot-dashed line at the right originating above the pivot shows a
 20\degree\ avoidance zone required to avoid accidental contact at
 launch.  The dashed parallelogram at the top in the right hand diagram,
 and to the left in the left hand diagram shows the shape of an
 extension to the Sun shields added for the Antarctic flight to allow
 observations to be made further from the anti-sun direction. }
	\label{fig:gondola}
\end{figure*}%

%%%%%%%%%%%%%%%%%%%%%%%%%%%%%%%%%%%%%%%%
\section{Gondola} \label{sec:gondola}
%%%%%%%%%%%%%%%%%%%%%%%%%%%%%%%%%%%%%%%%

The \blast\ gondola provides a pointed platform for the telescope and
the attachment point to the balloon flight train. The gondola was
designed and built by Empire Dynamic Dynamic
Structures,\footnote{\url{http://www.empireds.com/}} based on initial
designs from members of the \blast\ team.  Schematics of the gondola
are shown in Figures \ref{fig:gondola} and \ref{fig:gondola_dressed}.
The breakdown of the mass of various components is given in
Table~\ref{tab:mass}.

\subsection{Requirements}
The gondola design is driven by the pointing requirements of the
science case.  The elevation range of the inner frame, including the
2\,m primary mirror and the ${\sim}$\,200\,kg cryostat, is
25--60\degree. The entire gondola can rotate to any azimuth angle.
The in-flight pointing is accurate to ${\sim}$\,30\arcsec\ and
post-flight pointing reconstruction is accurate to better than
$5\arcsec$. The telescope's primary observing mode is to scan in
azimuth at ${\sim}$\,0.1\degree\,s$^{-1}$.  All of the mirrors are
shaded from solar radiation.

As a consequence of the pointing requirements, the feedback rate of
the control system is ${\sim}$\,10\,Hz. To accommodate this, the
gondola is designed to be rigid, with a minimum resonance frequency
greater than $14.4\,$Hz. All mechanical tolerances are set to minimize
backlash. All bearings are low friction, low stiction and treated with
low-temperature grease (see Section~\ref{SEC:ATTITUDECONTROLL}). The center-of-mass of the inner frame is on
the rotational axis (the elevation axis) so that translations of the
gondola (by wind or from the balloon) do not generate torques which
re-orient the telescope.

A target weight of ${\sim}$\,2000\,kg was set for the entire system
and the moment of inertia was estimated to be
${\sim}$\,4500\,kg\,m$^2$. The system has been engineered to meet the
flight criteria set by the Columbia Scientific Balloon Facility
(CSBF)\@.

\subsection{Layout}
The gondola frame consists of three components: an outer frame,
suspended from the balloon flight train by cables and a pivot motor;
an inner frame which is attached to the outer frame at two points
along a horizontal axis; and a set of Sun shields that attach to the
outer frame. The frame is constructed of light-weight aluminum tubing
and I-beam. \blast\ incorporates large-diameter torque motors for all
motion control.

\subsubsection{Major Components}
The outer frame consists of a horizontal surface, 2 yokes to support
the inner frame, and four legs. The outer frame is pointed in azimuth
using a reaction wheel and pivot.  It also provides mounts for various
electronics boxes, including the flight computers, the 
CSBF electronics,
various pointing sensors, and flight batteries.

The inner frame is made from thin-wall aluminum box beam. It supports the
telescope, cryostat, detector read-out electronics, gyroscopes and
star cameras. It is attached to the outer frame at two points,
defining an axis of rotation.

\begin{figure*}
        \centering
 	\includegraphics[width=\linewidth]{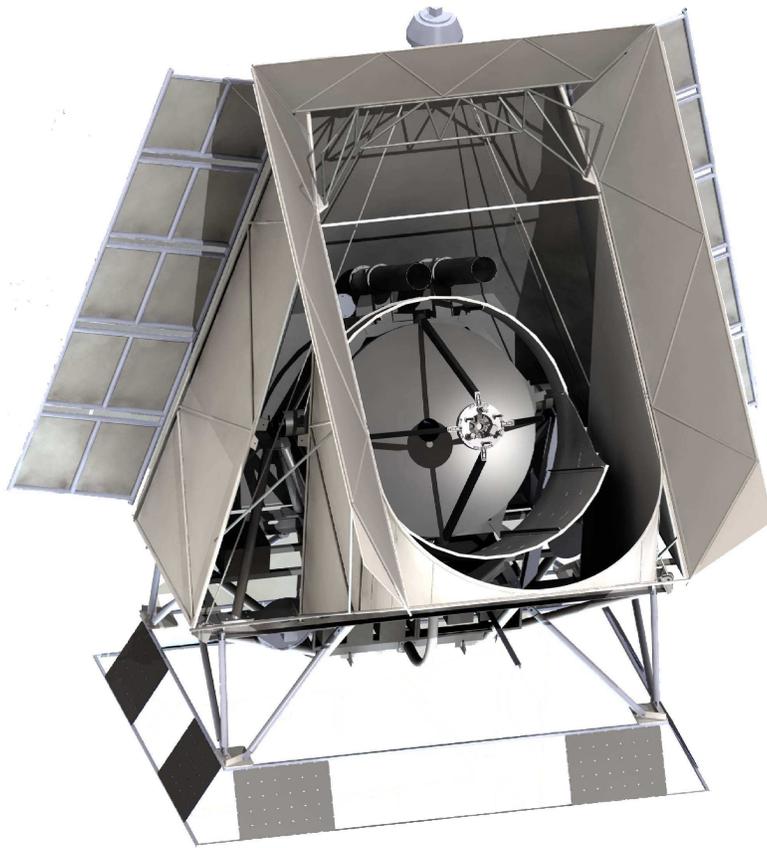}
 	\caption{A CAD model of the gondola, showing \blast\ fully
        assembled in the \blastmcm\ configuration. The electronics can cool
	radiatively through the triangular gaps
	which are visible in the Sun shields.  These gaps always point away
	from the Sun in flight.}
 	\label{fig:gondola_dressed}
 \end{figure*}

The back, sides and bottom of the gondola are surrounded by Sun
shields, allowing the telescope to point as low as 25\degree\ in
elevation. A lightweight aluminum frame encloses the telescope and is
covered with extruded polystyrene panels wrapped in aluminized Mylar. The original
design shielded the secondary mirror from the Sun at 30\degree\ above
the horizon, the highest the Sun rises during an Antarctic flight, with
the telescope pointing directly away from the Sun. Additional panels
were added for the \blastmcm\ flight to improve the shading when the
telescope is not pointed anti-Sun. The telescope is also protected
from solar radiation scattered from the ground. As well as the opening
for the telescope, there are forward-facing openings in the Sun shields
to allow for cooling of the electronics by radiation (see Figure~\ref{fig:gondola_dressed}).
\subsubsection{Attitude Control}\label{SEC:ATTITUDECONTROLL}
The telescope's attitude is controlled by three torque motors: the
reaction wheel, pivot and elevation motors. The motors are rare-earth,
permanent-magnet, direct-drive DC torque motors\footnote{Kollmorgen QT-6205.}
with a peak torque of 13.6\,N\,m.  They are controlled by 50\,A
pulse-width 
modulated servo amplifiers.\footnote{Advanced Motion Controls,
model 12A8K.}  The housings, made of hardened steel, are custom
designed and incorporate the bearings. The original design called for
a dry Teflon bearing lubricant, but we found that low-temperature
bearing grease\footnote{Dow
Corning Molykote\,$^{\mbox{\tiny\textregistered}}$ 33.}
 provides lower friction and stiction.

The telescope is controlled in azimuth by a 1.5\,m reaction wheel,
made of 7.6\,cm thick aluminum honeycomb,
and forty-eight 0.9\,kg brass disks mounted around the
perimeter to maximize the ratio of moment of inertia to mass. 
The reaction wheel is mounted at the center of the outer frame with its rotation
axis going through the pivot.
Torquing the gondola against the reaction wheel controls azimuth pointing.

\begin{figure*}
        \centering
 	\includegraphics[width=0.5\linewidth]{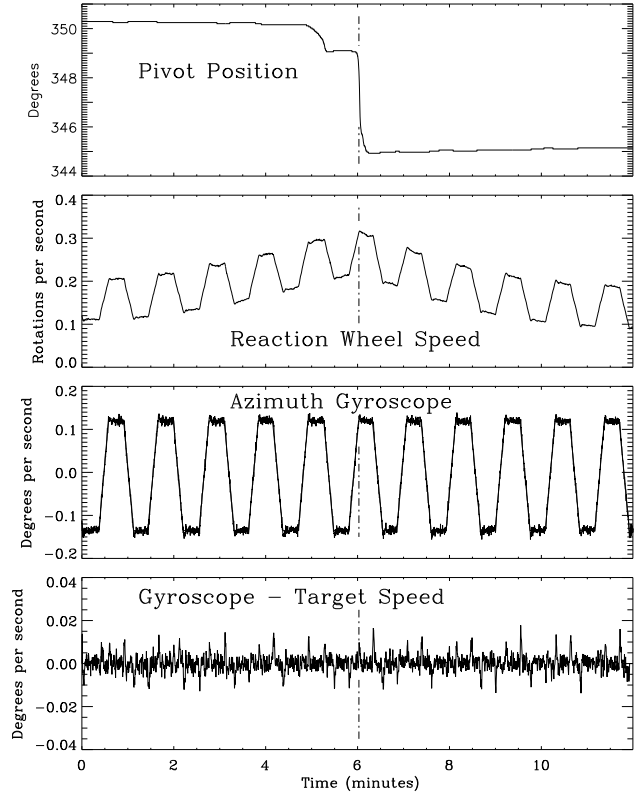}
\caption{Isolation of vertical torques.  All panels show data from the
same 12 minute segment of the 2006 flight, when the telescope was
executing a routine azimuthal scan.  The third panel from the top
shows a trapezoidal angular velocity profile corresponding to constant
angular velocity scans connected by constant angular accelerations.  
This panel also includes a plot of the desired angular speeds,
completely buried by the measured signal.   The second panel from the
top shows that azimuthal speed is governed by accelerating and
decelerating the speed of the reaction wheel.  A torque at the pivot
proportional to reaction wheel speed reaches threshold and overcomes
the stiction in the pivot when the reaction wheel speed is near 0.3
revolutions per second.  One such event is marked by the vertical
dashed line in all panels.  The bottom panel contains a plot of the
difference between measured gyroscope signal and desired azimuthal
speed.  Notice that there are small spikes visible corresponding to
the corners in the scan speed but that there is no azimuthal motion of
the telescope itself coincident with the release of stiction in the
pivot.}
\label{fig:motion}
\end{figure*}

The gondola is subject to external torque from wind shear acting on
the gondola  and from balloon rotation acting through the flight
train.   The second term is minimized, but not entirely eliminated,
 by a vertical axis pivot at the attachment  of the gondola to the
flight train (see Figure~\ref{fig:gondola}).  Since the flywheel motor
saturates at ${\sim}$\,50\,rpm, the additional torque motor at the
pivot is needed to reduce excess angular momentum, keeping the
rotation of the reaction wheel within reasonable limits.  
Figure~\ref{fig:motion} demonstrates that this nested servo system,
where the reaction wheel is driven to maintain telescope orientation
and the pivot is driven to maintain low reaction wheel speed,
successfully decouples the telescope from external torques.

The elevation of the inner frame is controlled by a motor mounted on
one side of the inner frame at the attachment point to the outer
frame.  A free bearing provides the connection point on the other
side. The motor housing incorporates a 16-bit shaft
encoder\footnote{Gurley Precision Instruments A25SB.}  which measures
the orientation of the inner frame relative to the outer frame. A
fluid transfer system, consisting of a pump and two tanks containing
${\sim}$\,10\,L of a high-density non-toxic fluid,\footnote{Dynalene
HC-40.} corrects for long time-scale balance offsets induced by the
boil-off of the cryogens and to the redistribution of cryogens due to
changes in elevation.

The balloon environment introduces small oscillations with a period of
approximately 25\,s in the pitch and roll of the gondola. Typical
amplitudes of ${\sim}$\,2\arcmin\ are observed. A roll damping system
consisting of a motor and a small 30\,cm diameter reaction wheel is
mounted perpendicular to the flywheel on the outer frame, with the
goal of dissipating roll modes.  The reaction wheel is made of
aluminum and weighs ${\sim}$\,2\,kg, most of which is at the
perimeter. Oscillations in pitch are removed by controlling the
elevation of the inner frame.  The oscillations in pitch and roll
where measured to be 50\% smaller compared to a case when both
controllers were off.

%%%%%%%%%%%%%%%%%%%%%%%%%%%%%%%%%%%%%%%%%%%%%%
\section{Command and Control} \label{sec:control}
%%%%%%%%%%%%%%%%%%%%%%%%%%%%%%%%%%%%%%%%%%%%%

\blast\ is designed to operate autonomously without need of ground
commanding.  Telescope control is provided by a pair of redundant
flight computers with Intel Celeron processors at 366 and 850\,MHz,
running Slackware\footnote{\url{http://www.slackware.com/}} Linux 9.2
with Kernel version 2.6.8\@. The computers are kept at near
atmospheric pressure to allow the hard drives to function properly and
to provide the appropriate thermal environment for the CPUs.

The flight computers run a single, monolithic program, the ``master control
program'' (mcp), written in C,  which performs primary control of all aspects of the
telescope including in--flight pointing solution,  motion, commanding,
telemetry, data archiving, and thermal control.  

The communication link between the telescope and the ground is
provided by CSBF through a number of line-of-sight (LOS) transmitters and
satellite links. The ground station computers interface with CSBF's ground station equipment to send commands to the payload and display
down-linked data. The ground station software uses
kst\footnote{\url{http://kst.kde.org/}} and other Linux applications
developed by \blast\ team members.

\subsection{Gondola Electronics}
The flight computers are built upon a passive PCI-bus backplane and
contain a single board computer, a four port 8250 serial port
extender card, and a custom made PCI controller card, which
interfaces major components as well as the LOS data
transmitter. This card interfaces with the
BLASTbus, a proprietary RS485 bus with three differential pairs (data,
clock, strobe).  The BLASTbus protocol is a poll and response
architecture composed of a 32-bit request followed by a 32-bit
response containing 16 bits of data.  The BLASTbus clock runs at 4\,MHz,
giving an effective bandwidth on the BLASTbus of 1\,Mbit\,s$^{-1}$.

The BLASTbus connects the flight computers to the Attitude Control
System (ACS) and DAS.  Each DSP card in these systems 
monitors the
BLASTbus. They provide data on request and receive command input
from the flight computers via the BLASTbus.

Data are marshaled on the BLASTbus into 100\,Hz frames.  Bolometers are
polled once per frame, as are the gyroscopes and other high-precision
pointing data.  The majority of housekeeping signals do not need to be
polled at 100\,Hz; these ``slow data'' signals are polled at 5\,Hz with
groups of twenty multiplexed into a single 100\,Hz channel.  The 100\,Hz
frames provide the basic data structure for data archival. The frames
are written to disk and transmitted by the line-of-sight 1\,Mbit\,s$^{-1}$ 
transmitter.

In addition to the BLASTbus, the flight computers also collect data
via 8250-style serial connections to the differential GPS, the SIP
computers, the lock motor, and the secondary actuators.  Ethernet
provides further connectivity to the star camera computers and Sun
sensor computer.  In order to provide data synchronization, the
auxiliary channels are written back to the BLASTbus and the star
cameras are triggered via the BLASTbus.

\subsection{Pointing Control}
\blast\ is a scanning experiment.  The primary scan-mode involves an
azimuthal raster coupled with a slow elevation drift or discrete
elevation steps.  

Detector response times, $\tau$, together with $1/f$ noise and star camera
integration times set limits on the azimuthal scan rate, $v_{{\rm az}}$.  
Two samples per FWHM are required to fully sample a Gaussian beam.
The maximum angular rate allowed by the detector $\tau$
is therefore
\[
	v_{{\rm az}} = \frac{\mbox{\scriptsize FWHM}/2}{2\pi\,\tau} 
\]
or about 0.3\degree\,s$^{-1}$ for the 30\arcsec\ beam at 250\,\micron\ 
and $\tau\sim 2$\,ms.  Mid-scan smearing in the star cameras also
limits azimuthal scan speeds to about 0.1\degree\,s$^{-1}$. A scan
rate larger than this is acceptable only for relatively small scans
where star camera solutions at the turn-around are more frequent. Low
frequency ($1/f)$ noise sets the largest angular mode that can be
constrained to $v_{{\rm az}}/f_0 \lesssim 3$\degree\ at
0.1\degree\,s$^{-1}$ ($f_0$ is the $1/f$ knee at 35\,mHz).
Therefore, \blast\ scans most of the time at $v_{{\rm az}} \simeq
0.1$\degree\,s$^{-1}$, occasionally going twice as fast on small
($\lesssim 0.3$\degree\ wide) maps.  Azimuthal acceleration is limited
to 0.1\degree\,s$^{-2}$ by the control electronics, resulting in,
typically, 2\,s for an azimuthal turn-around.

Elevation drift speeds of 10\arcsec\,s$^{-1}$ or elevation steps of
40--100\arcsec\ in adjacent azimuth scans provide adequate
spatial sampling of the sky by the detector arrays.

\begin{figure*}
        \centering
       	\includegraphics[angle=-90, keepaspectratio, width=\linewidth]{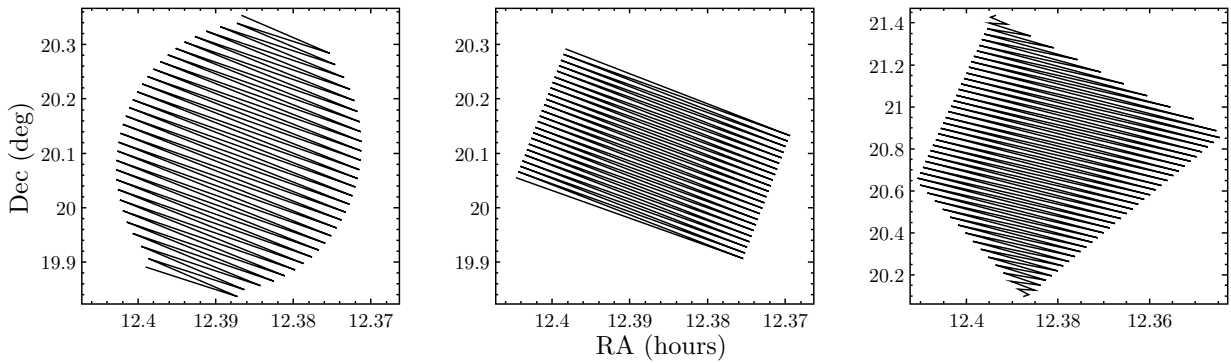}
	\caption{Idealized representations of \blast's three scan modes: from left to right a ``cap'', a ``box'', and a ``quad''. In practice, the gyroscopes are used to control the speed and orientation of each single scan, while star camera--based orientation solutions at the endpoints are used to prevent gyroscope errors from accumulating.  }
	\label{fig:scan_modes}
\end{figure*}

Three primary scan modes are implemented (Figure~\ref{fig:scan_modes}): ``cap'', a circle centered on a
target RA/Dec; ``box'', a rectangle in azimuth and elevation centered on a
target RA/Dec; and ``quad'', an arbitrary quadrilateral specified by its
four corners in RA and Dec. 

%%%%%%%%%%%%%%%%%%%%%%%%%%%%%%%%%%%%%%%%%%%%%%%%%%%%%%%
\section{Pointing Sensors} \label{sec:psens}
%%%%%%%%%%%%%%%%%%%%%%%%%%%%%%%%%%%%%%%%%%%%%%%%%%%%%%%

The primary pointing sensors for \blast\ are a pair of CCD-based star
cameras and Fiber-Optic rate gyroscopes. The star cameras provide
absolute pointing and the gyros provide velocity information which can
be integrated to allow interpolation of the gondola's attitude between
star camera solutions. Coarse attitude determination is provided by
several additional sensors: in elevation, there is an encoder on the
elevation axis and a tilt sensor on the inner frame; in azimuth, there
is a Sun sensor, a differential GPS unit, and a magnetometer.  The
system provides post-flight pointing reconstruction to $\lesssim
5\arcsec$\ rms.

\subsection{Star Cameras}

Four primary factors drive the design of the star cameras: (i) an
absolute pointing accuracy of ${\sim}$\,5\arcsec\ is required to
over-sample the diffraction-limited size of the 250\,\micron\ beam;
(ii) integration times (and hence efficiency) have to be short enough
to avoid significant smearing at the maximum normal scan
angular-velocity of the gondola (0.1\degree\,s$^{-1}$); (iii) the
system must {\em always\/} detect stars to calibrate gyroscope drift;
and (iv) the frequency of the solutions must be high enough to control
the $1/f$ random walk noise in the integrated gyroscopes.

To meet these requirements, we incorporated two star cameras for
redundancy, and to enable increased positional accuracy in post-flight
processing.  A detailed description of the cameras can be found
in~\citet{rex06}.
The cameras each use a Nikon lens with a 200\,mm focal length and a
100\,mm aperture to produce a 2.5\degree\ $\times$ 2.0\degree\ FOV
with 7\arcsec\ pixels.  Each camera is controlled by its own PC/104,
300\,MHz Celeron computer which calculates pointing solutions for the
gondola at a rate of ${\sim}$\,1.5\,Hz.  The computers command the CCD
cameras via FireWire, control the focus and aperture size using
stepper motors via a serial port, and regulate the temperature of the
camera using a small USB DAQ module.  The entire system is contained
in a pressure vessel to allow the operation of the hard drives,
control the thermal environment, and maintain mechanical rigidity.
Since the cameras operate during the day, the dominant source of noise
is from the sky background, despite the altitude of the experiment. A
Nikon R60 red filter is used to attenuate the background. In addition,
a 1.2\,m long cylindrical baffle is attached to the front of each
camera to reduce stray-light contamination beyond 10\degree\ from the
optical axis.

The software controlling the camera in flight provides real-time
pointing information by analyzing the star patterns in the CCD frames.
The pointing algorithm locates blobs in the camera image, rejecting
the known bad pixels.  The best fit positions of star candidates are
then used by a pattern recognition algorithm to identify a unique
constellation matching the observed angular separations in a star
catalog \citep[Guide Star Catalog 1.1, see][]{lasker87}.  The
magnitude limit of the catalog is chosen manually and no brightness
information for the stars is otherwise used. The algorithm is aided by
an approximate pointing solution from the flight computer, required to
be accurate to ${\sim}$\,5\degree\ (see Section~\ref{sec:CS}) to
reduce the number of candidate star identifications. A visual
magnitude limit of 9 was required to obtain sufficient completeness. A
``Lost in Space'' algorithm based on the Pyramid technique
\citep{mort01} was also implemented to be used in case the approximate
solution was found to be unreliable (which never happened during the
\blast\ flights).

Once the CCD object centroids are matched to stars with known
coordinates $(\alpha,\delta)$, the pointing solution is calculated,
parametrized by the celestial coordinates of the center pixel,
$\alpha_0$ and $\delta_0$, and the roll of the camera, $r$. A model in
which the image is assumed to be a perfect gnomic tangent-plane
projection, with the tangent point at $\alpha_0$ and $\delta_0$, is
used to project each star RA and Dec into the plane of the CCD.  The
rms distance between the CCD and model star coordinates is then
minimized using an iterative Newton solver with respect to the three
model parameters. This procedure produces pointing solutions with
uncertainties of ${\sim}$\,3.5\arcsec\ and ${\sim}$\,200\arcsec\ for
the position of the tangent point and of the roll, respectively.  A
post-flight comparison of simultaneous pointing solutions from both
cameras results in an rms uncertainty of $<$\,2\arcsec.

\subsection{Gyroscopes}
Fiber-Optic rate Gyroscopes (FOG) are used to extrapolate the star
camera absolute attitude to provide pointing information at each
detector time sample. Two redundant sets of three FOGs are
mechanically arranged to measure the gondola angular velocity along
orthogonal axes. One of these sets uses ECORE 2000 analog gyroscopes
while the other uses DSP 3000 digital output gyroscopes from the same
company.\footnote{KVH Industries, Inc.} They have an angle random walk
of 5\arcsec\,s$^{-1/2}$ and 4\arcsec\,s$^{-1/2}$, respectively.  Since
FOGs are sensitive to magnetic fields \citep{bohm}, they were wrapped
in a 250\,\micron\ thick $\mu$-metal sheet. This reduced the signal
induced from the Earth's magnetic field by a factor ${\sim}$\,10. The
gyroscope assembly is temperature controlled to mitigate bias drifts.
For \blastkir\, both sets worked flawlessly.  In \blastmcm\ two out of
the three digital gyroscopes stopped working temporarily, possibly due
to a cosmic ray interaction with the gyro electronics. A power cycle
of the gyroscope itself restored function.  Post-flight attitude
reconstruction of the 2005 and 2006 data shows that the FOGs performed
at near to their specified sensitivity.

\subsection{Coarse Sensors} \label{sec:CS}
Two distinct Sun sensors were used on \blast\ for coarse azimuth determination.
For \blastfs\ and \blastkir, a linear CCD behind a thin slit was used.  It was replaced by an array of photo-diodes
arranged in a azimuthally oriented ring for \blastmcm.  
The photo-diode Sun sensor was smaller, lighter, and consumed less power 
(5\,W compared to 50\,W) than the CCD-based sensor.  

The photo-diode Sun sensor is a small, all-in-one unit.  A PC/104
embedded computer system is mounted below a dodecagonal ring.  Twelve
photo-diodes are mounted radially around the ring, and the ring is
mounted with its axis parallel to the azimuth axis.  The intensities
of light incident on the photo-diode with the highest intensity and on
its 2 nearest-neighbors are compared and fit to a $\cos \theta$
function to determine the azimuth relative to the Sun.  The flight
computer calculates the azimuth of the Sun to determine the azimuth of
the gondola.  The CCD Sun sensor, used during the first science
flight, had a short-time, relative precision of 2\arcmin, but an
overall absolute accuracy of 5\degree\@.  The photo-diode Sun sensor,
used during the second science flight, had a precision of 4\arcmin,
with an absolute accuracy of 5\degree\@.  Both units provide azimuthal
data at $>5$\,Hz.

\blast's second coarse azimuth sensor was an ADU5 differential GPS
unit.\footnote{Magellan Navigation, Inc.}  The four required antennae
were installed on booms above the Sun shields to minimize multi-path
reflections from the gondola (see Figure~\ref{fig:gondola}).  This
orientation was predicted to provide better than 10\arcmin\ rms
absolute pointing.  However, for \blastkir, the GPS stopped providing
azimuthal solutions before the instrument reached float altitude and
never recovered.  For \blastmcm, the GPS did not provide solutions
until 110 hours after launch.  It then began providing solutions and
the unit operated with a precision from 0.1\degree\ to~0.2\degree\
rms.  The cause of the irregular reliability of the GPS is uncertain,
but may be due to a lack of thermal rigidity in the antenna mounts.
The position, altitude, velocity, and time data provided by the GPS
unit worked well at all times.

\blast's third and most reliable coarse azimuth sensor was a 3-axis,
flux-gate magnetometer\footnote{Applied Physics, model 534.} which was
used to determine the gondola's attitude relative to the Earth's
magnetic field.  The magnetic field orientation was determined using
the World Magnetic Model.\footnote{WMM-2005, National
Geospatial-Intelligence Agency (NGA).}  Even though \blast\ passed
close to the magnetic pole, the model was accurate enough and the
gondola pendulations small enough that the magnetometer-based pointing
solution was good to 5\degree\ peak to peak in both science flights.

\section{Thermal Environment} \label{sec:thermal}
The LDB thermal environment is characterized by continuous but
variable solar illumination combined with small coupling to the
atmosphere.  Extensive shielding is used to regulate the temperature
of the instrument.  Following the design of BOOMERANG \citep{crill03},
\blast\ is surrounded by shields comprised of 24\,\micron\ thick
aluminized Mylar applied to 2.5\,cm thick polystyrene open-celled
foam.  The Mylar side faces out on all surfaces which may be exposed
to the Sun.  While bare aluminum surfaces can be expected to reach
temperatures in excess of $130^{\circ}$C in the LDB float environment,
Sun-facing aluminized Mylar surfaces reach temperatures of around
$45^{\circ}$C due to their high ratio of thermal IR emissivity to
visible light absorptivity.  For telescope azimuth ranges of
$\pm$60\degree\ from anti-Sun, the shielding blocks direct radiation
from reaching the optics and all electronics, except for the pivot
motor controller and Sun sensor. The actual angles of Sun avoidance
have a weak dependence on the elevation of the telescope.

The majority of the electronics dissipate their heat radiatively.  A
high IR emissivity is achieved by painting electronics enclosures with
white paint.\footnote{We used a variety of generic spray-on epoxy
enamel.}  Painting the interiors of boxes white aids in radiative
transfer between the electronics and the box.  This is effective for
all boxes, except for the detector readout electronics boxes, each of
which dissipates about 100\,W.  To cool these boxes, emissive vertical
plates were installed between each card to couple heat from the
electronics to the top and bottom walls of the boxes.  A closed-cycle
fluid cooling loop was used to distribute the heat to the gondola
frame.

The thermal strategy was effective in keeping the temperatures of
essentially all electronics well within their operational range.  A
shield added to \blastmcm\ to keep the Sun from illuminating the
secondary mirror reflected solar radiation back onto the pivot motor
controller.  This increased the average temperature of the pivot by up
to $20^{\circ}$C compared to what was experienced in \blastkir\@.  As
a result, two non-essential observations where shortened when the
motor controller temperature exceeded $62.5^{\circ}$C.

For \blastmcm, the primary mirror was thermally isolated from the
inner frame with G10 spacers. Approximately twelve layers of Mylar
Super-insulation\footnote{Cryolam\,$^{\mbox{\tiny\textregistered}}$}
were added in the gap between the mirror and the frame. This
stabilized the in-flight temperature, which exhibited a diurnal
excursion of about $\pm$1.5\degree\/C\@.

\begin{figure*}
        \centering
	\includegraphics[angle=-90, width=\linewidth]{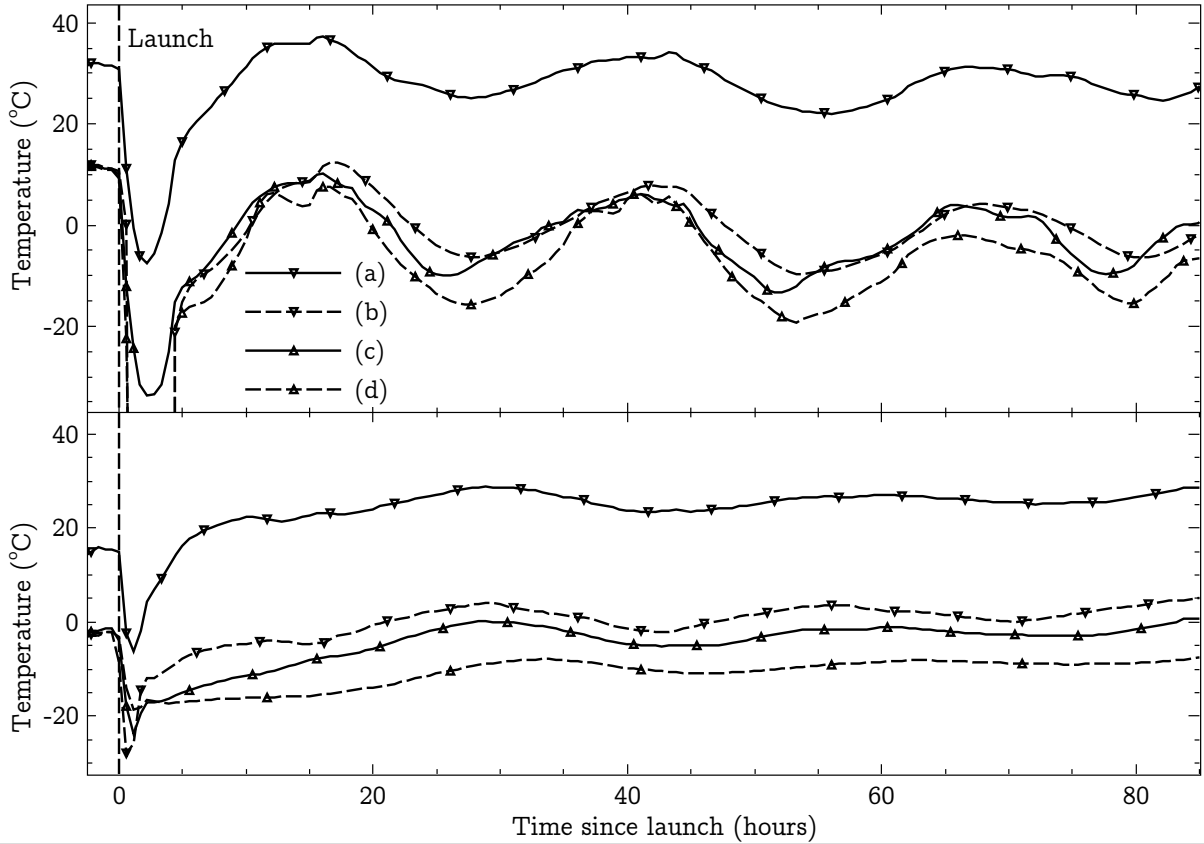}
	\caption{ Temperatures during the 2005 (top panel) and 2006
(bottom) flights.  Several gondola temperatures are plotted for the
period starting just before launch and extending through the first few
days from both science flights.  Dramatic cooling in the troposphere,
reheating in the stratosphere and diurnal variations due to variation
of the solar elevation angle are visible in all traces.  The curves
are a: the pressure vessel containing the data acquisition computers,
b: the gondola inner frame, c: the elevation axis motor and d: the
primary mirror.  The temperature difference between high power
consumption electronics (a) and more passive components (b and c) is
only slightly larger at float than it is on the ground.  The primary
mirror has the best geometry for cooling radiatively and therefore
attains the coldest temperature.  Notice the relatively large
variations in the primary mirror temperature in the 2005 flight which
caused us to recognize the need for in-flight focussing capability
combined with better insulation of the primary.}
\label{fig:thermal}
\end{figure*}

See Table~\ref{tab:thermal} and Figure~\ref{fig:thermal} for a list of temperatures achieved during the various \blast\ flights.%

\section{Power System}
Solar panels provide power to the flight electronics.  They are
mounted at the back of the telescope on the support structure for the
Sun shields (see Figure \ref{fig:gondola}). The panels face the Sun
from only one side and are radiatively cooled to the sky from the
other side.

During normal operation, \blast\ requires 500\,W, divided between two
completely isolated systems: the receiver, and the rest of the gondola
(see Table~\ref{table:power}). During line-of-sight operations, an
additional 150\,W are required to power the transmitters. Power is
provided by the solar power system\footnote{MEER Instruments, Palomar
Mountain, CA, http://www.meer.com/ and SunCat Solar, Phoenix, AZ.}
which charges NiMH batteries.\footnote{Cobasys Model 9500.} The arrays
provide 1250\,W for normal incidence at float. The arrays are thus
able to provide full power to the experiment up to a Sun-array angle
of $>60^{\circ}$. The batteries are essentially only used during
launch and ascent.

As a NiMH battery reaches its top-off voltage, its charge efficiency
decreases, leading to heating of the battery. Additionally, the topoff
voltage of NiMH batteries decreases with increased temperature. This
combination can lead to thermal runaway if the batteries are charged
with a fixed top-off voltage.  For \blast, the charge controllers were
modified to allow the fight computer to reduce the top-off voltage as
the battery temperature increases, permitting full battery charge
without thermal runaway.

%%%%%%%%%%%%%%%%%%%%%%%%%%%%%%%%%%%%%%%%%%%%%%%
\section{Flight Planning}

\blast\ functions autonomously using schedule files that consist of a
sequential list of observations or actions as a function of the local
sidereal time.  This system is robust against temporary system
failures because the telescope only needs to know the current time and
location to resume operation upon recovery.  Using a local sidereal
clock rather than a clock fixed in some time zone, it is possible to
account for purely astronomical visibility constraints (such as the RA
of the Sun and of the astronomical targets) using a static
description. However, the amount of time the telescope spends on a
given observation varies depending on telescope longitudinal drifts.

Flight schedules are calculated on the ground, in advance of the
launch using best-estimates for the launch time, flight path and
length.  Given the uncertainties in these flight parameters and the
complexity of the observations (many small maps scattered across the
sky, of which only a subset are available at a given instant), a
schedule file generator was developed to automate the process.  This
system interprets a high-level description of the desired scientific
observations, such as ``map a circular region centered at a given RA
and DEC with a given radius''.  It then accounts for a number of
parameters such as: the launch date and time, projected duration of
flight, the start longitude and latitude, a guess for the termination
longitude, a latitude range over which the payload may drift, and the
position of the Sun and the Moon.  With this information, the scheduler
generates an optimized list of consecutive actions.  At each instant,
observations that are possible under the current visibility
constraints are scheduled in order of scientific priority.  At regular
intervals, calibration observations are assigned maximum priority, so
that sensitivity, beam shape, and pointing variations may be tracked
throughout the flight.  Once a schedule file is generated, a simple
model for the power usage of the system and the charge rate of the
solar panels as a function of Sun incidence angle is used to
calculate the available battery charge over time. The schedule is
modified by hand if this model indicates that the batteries will have
less than 50\% of the total charge capacity at any point in the
flight.  Schedules are then tested using a simulator that mimics the
scanning motion of the telescope to determine the coverage (effective
integration time across the maps) and cross-linking among each of the
maps for a given flight trajectory.

Six schedule files are generated for each launch opportunity.  The
latitude of the gondola can change by as much as ${\sim}$\,15\degree\
during a flight, hence three different schedules are created in
latitude bands that are 6\degree\ wide, with 1\degree\ of overlap.
The gondola uses the time and payload position reported by the GPS to
decide which schedule file to use.  Two sets of these three schedules
are made: the first set assumes the instrument is working with the
target sensitivities; the second assumes degradations of the telescope
beam size by a factor of $\sqrt{2}$, and sensitivity by a factor of 2.
At the beginning of the flight, the sensitivities are estimated from
scans across calibrators.  Based on this information the ground
station team can decide which of the two sets of schedule files the
instrument should use, and switch between the two using a single
command.

\section{Pointing Reconstruction} \label{sec:performance}
\begin{figure*}
        \centering
	\includegraphics[angle=-90, width=\linewidth]{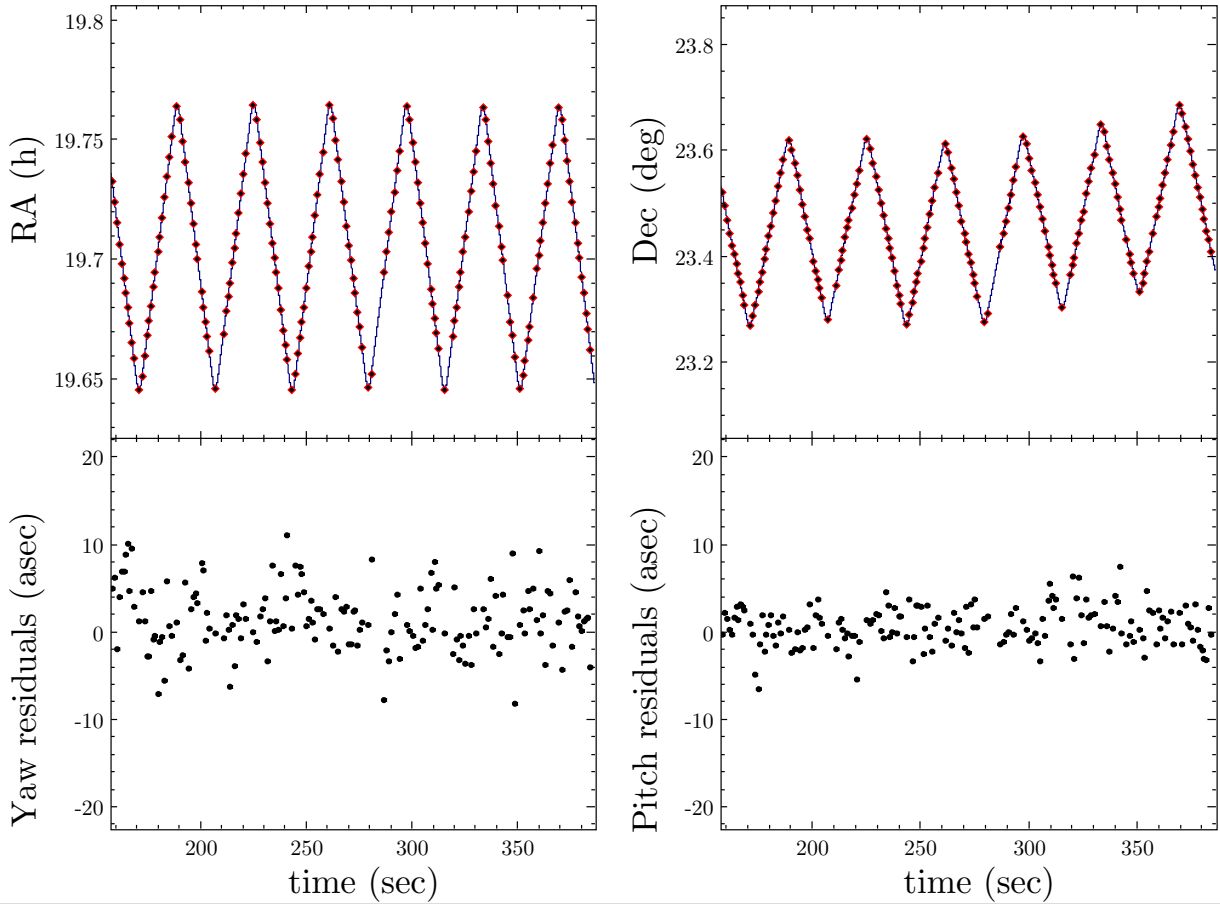}
	\caption{An example of pointing reconstruction from the
	\blastkir\ data. In the upper panels, the solid blue lines
	represent the reconstructed pointing solution obtained by
	integrating one set of gyroscopes onto one of the two star
	cameras.  Dots represent the positions reported by the other
	star camera, which is not used in the solution. The lower
	panels show the residuals as yaw (${\sim}\,\Delta {\rm
	RA}\times\cos{\rm (Dec)}$) and pitch (coincident with Dec in
	this particular case).  The yaw residuals are slightly wider,
	since the telescope scans in azimuth, which is close to RA at
	the \blastkir\ latitudes. The projection of the gondola
	angular velocity along the yaw axis is always larger than the
	pitch axis for the elevation range of the telescope
	(25\degree--60\degree), and hence the star centroids in star
	camera images have larger measurement errors in that
	direction.  The overall error of the pointing solution can be
	obtained by summing in quadrature the standard deviations of
	the residuals, yielding $\simeq 4.2$\arcsec\ rms. Improvements
	to this result are discussed in Section~\ref{sec:performance}}
	\label{fig:psol_a}.
\end{figure*}

Post-flight pointing reconstruction estimates the rotation (attitude)
of the gondola with respect to the celestial sphere as a function of
time, providing RA, Dec, and rotation angle information at each
detector sample.  The star camera provides absolute attitude on an
unevenly sampled time grid (${\sim}$\,1.5\,s), with an accuracy of $<
2$\arcsec\ rms.  Each solution is sampled at a known phase with
respect to the detectors.  The detector and gyroscope sampling are
synchronized; therefore, the integration of the gyroscopes gives one
estimate of the gondola attitude. The star camera is used to correct
the random walk drift induced by the integrated gyroscope noise and to
give an estimate of the integration constant.  \blast\ implements a
Kalman filter approach to estimate the attitude, expressed as the
state quaternion $q(t)$, a 4-dimensional quantity describing the solid
body rotation of the gondola.  The non-linear state model representing
the gondola is defined as
\begin{equation}
        \left\{\begin{array}{l}
        \vec{b}_{n+1} = \vec{b}_n + {\vec {w_b}}_n\\
        q_{n+1} =  \qvec{1}{\frac{1}{2}\left(\vec\omega_n +  \vec{b}_n + \vec{w_\omega}_n \right) {\scriptstyle\Delta T}}q_n\\
        \end{array}\right.
\end{equation}
where $\vec\omega_n$ is the gyro angular velocity with
$\vec{w_\omega}_n$ its noise, $\vec{b}_n$ is the gyro bias,
$\vec{w_b}_n$ is the filter process noise, and $\Delta T$ is the time
resolution of the system (10\,ms).  The filter is run forward and
backward in time and the two solutions are weighted together, using
the Kalman covariance matrix output from the filter as the
weight. Using just one star camera and the digital gyros, the final
averaged attitude is better than $5$\arcsec\ rms; one example is
shown in Figure~\ref{fig:psol_a} using data from
the \blastkir\ flight. The achieved precision is sufficient for \blast\
beam sizes, but can be improved once the second set of gyroscopes and the additional 
star camera
are included in the solution. 
An improvement of $\sqrt{2}$ to 2 better can
be reasonably expected.
The pointing solution is referenced to the star camera reference frame and needs
to be rotated into the submillimeter array coordinate frame. 
Bright optical and submillimeter point sources
are used to evaluate the rotation quaternion $Q$. 
This correction is ideally static, but, in
practice, was found to be a weak function of time. As shown in
Figure~\ref{fig:gondola}, the star cameras are mounted at the top of
the inner frame. A  structural deformation of the frame can modify the
relative orientation between the cameras and the submillimeter beam.
Therefore, several pointing calibrators are observed each time the
geometry of the scan changes. This effect can be as large as
35\arcsec\ and has an 80\% correlation with the elevation, and a 25\%
correlation with the temperature of inner frame. Although it is
possible to model the effect based on the gondola thermometry and
attitude, $Q$ is such a weak function of time that this was found unnecessary.

\section{Conclusion}

\blast\ is an instrument designed to probe the
local and high-z Universe at short submillimeter wavelengths.  Over the course of two
science flights, \blast\ has addressed a broad range of Galactic and
extragalactic topics. The balloon environment requires a novel combination
of ground and satellite technology and techniques.  This paper is
intended to serve as a reference for a suite of science papers from
the \blastkir\ and \blastmcm\ flights.
\\

The \blast\ collaboration acknowledges the support of
NASA through grant numbers NAG5-12785, NAG5-13301 and NNGO-6GI11G, the
Canadian Space Agency (CSA), the Science and Technology Facilities Council (STFC),
Canada's Natural Sciences and Engineering Research Council (NSERC),
the Canada Foundation for Innovation,
the Ontario Innovation Trust,
the Puerto Rico Space Grant Consortium, the Fondo Istitucional
para la Investigacion of the University of Puerto Rico, 
and the National Science Foundation Office of Polar Programs;
C.~B. Netterfield also acknowledges support from the Canadian Institute for Advanced Research. 
L.~Olmi would like to acknowledge Pietro Bolli for his help with Physical 
Optics simulations during the testing phase of the \blastmcm\ telescope.
We would also like to thank the Columbia Scientific Balloon Facility
(CSBF) staff for their outstanding work, the Precision Machining Group at
Lawrence Livermore Laboratory, the support received from Empire Dynamic Structures in the design and construction of the gondola,
Daniele Mortari for helpful discussions in the development of the Pyramid code, 
Dan Swetz for buliding the Fourier transform spectrometer,
and  Luke Bruneaux, Kyle Lepage,  Danica Marsden, Vjera Miovic, and James Watt for their contribution to the project.

%%%%%%%%%%%%%%%%%%%%%%%%%%%%%%%%%%%%%

\bibliography{ms}

\begin{thebibliography}{26}
\expandafter\ifx\csname natexlab\endcsname\relax\def\natexlab#1{#1}\fi

\bibitem[{{Ade} {et~al.}(2006){Ade}, {Pisano}, {Tucker}, \& {Weaver}}]{ade06}
{Ade}, P.~A.~R., {Pisano}, G., {Tucker}, C., \& {Weaver}, S. 2006, in Presented
  at the Society of Photo-Optical Instrumentation Engineers (SPIE) Conference,
  Vol. 6275, Millimeter and Submillimeter Detectors and Instrumentation for
  Astronomy III. Edited by Zmuidzinas, Jonas; Holland, Wayne S.; Withington,
  Stafford; Duncan, William D.. Proceedings of the SPIE, Volume 6275, pp.
  62750U (2006).

\bibitem[{{Bock} {et~al.}(1996){Bock}, {Delcastillo}, {Turner}, {Beeman},
  {Lange}, \& {Mauskopf}}]{bock96}
{Bock}, J.~J., {Delcastillo}, H.~M., {Turner}, A.~D., {Beeman}, J.~W., {Lange},
  A.~E., \& {Mauskopf}, P.~D. 1996, in ESA SP-388: Submillimetre and
  Far-Infrared Space Instrumentation, ed. E.~J. {Rolfe} \& G.~{Pilbratt}, 119

\bibitem[{{Bock} {et~al.}(1998){Bock}, {Glenn}, {Grannan}, {Irwin}, {Lange},
  {Leduc}, \& {Turner}}]{bock98}
{Bock}, J.~J., {Glenn}, J., {Grannan}, S.~M., {Irwin}, K.~D., {Lange}, A.~E.,
  {Leduc}, H.~G., \& {Turner}, A.~D. 1998, in Presented at the Society of
  Photo-Optical Instrumentation Engineers (SPIE) Conference, Vol. 3357, Proc.
  SPIE Vol. 3357, p. 297-304, Advanced Technology MMW, Radio, and Terahertz
  Telescopes, Thomas G. Phillips; Ed., ed. T.~G. {Phillips}, 297--304

\bibitem[{{Bohm} {et~al.}(1982){Bohm}, {Petermann}, \& {Weidel}}]{bohm}
{Bohm}, K., {Petermann}, K., \& {Weidel}, E. 1982, Optics Letters, 7, 180

\bibitem[{{Chapin} {et~al.}(2008)}]{Chapin2008}
{Chapin}, E. {et~al.} 2008, ApJ, accepted

\bibitem[{Chattopadhyay {et~al.}(2003)Chattopadhyay, Glenn, Bock, Rownd,
  Caldwell, \& 2003}]{chatt03}
Chattopadhyay, G., Glenn, J., Bock, J.~J., Rownd, B.~K., Caldwell, M., \& 2003,
  M. J.~G. 2003, IEEE Trans. Micro. T. Tech

\bibitem[{{Crill} {et~al.}(2003){Crill}, {Ade}, {Artusa}, {Bhatia}, {Bock},
  {Boscaleri}, {Cardoni}, {Church}, {Coble}, {de Bernardis}, {de Troia},
  {Farese}, {Ganga}, {Giacometti}, {Haynes}, {Hivon}, {Hristov}, {Iacoangeli},
  {Jones}, {Lange}, {Martinis}, {Masi}, {Mason}, {Mauskopf}, {Miglio},
  {Montroy}, {Netterfield}, {Paine}, {Pascale}, {Piacentini}, {Polenta},
  {Pongetti}, {Romeo}, {Ruhl}, {Scaramuzzi}, {Sforna}, \& {Turner}}]{crill03}
{Crill}, B.~P., {Ade}, P.~A.~R., {Artusa}, D.~R., {Bhatia}, R.~S., {Bock},
  J.~J., {Boscaleri}, A., {Cardoni}, P., {Church}, S.~E., {Coble}, K., {de
  Bernardis}, P., {de Troia}, G., {Farese}, P., {Ganga}, K.~M., {Giacometti},
  M., {Haynes}, C.~V., {Hivon}, E., {Hristov}, V.~V., {Iacoangeli}, A.,
  {Jones}, W.~C., {Lange}, A.~E., {Martinis}, L., {Masi}, S., {Mason}, P.~V.,
  {Mauskopf}, P.~D., {Miglio}, L., {Montroy}, T., {Netterfield}, C.~B.,
  {Paine}, C.~G., {Pascale}, E., {Piacentini}, F., {Polenta}, G., {Pongetti},
  F., {Romeo}, G., {Ruhl}, J.~E., {Scaramuzzi}, F., {Sforna}, D., \& {Turner},
  A.~D. 2003, \apjs, 148, 527

\bibitem[{{Devlin} {et~al.}(2004){Devlin}, {Ade}, {Aretxaga}, {Bock}, {Chung},
  {Chapin}, {Dicker}, {Griffin}, {Gundersen}, {Halpern}, {Hargrave}, {Hughes},
  {Klein}, {Marsden}, {Martin}, {Mauskopf}, {Netterfield}, {Olmi}, {Pascale},
  {Rex}, {Scott}, {Semisch}, {Truch}, {Tucker}, {Tucker}, {Turner}, \&
  {Weibe}}]{devlin04}
{Devlin}, M.~J., {Ade}, P.~A.~R., {Aretxaga}, I., {Bock}, J.~J., {Chung}, J.,
  {Chapin}, E., {Dicker}, S.~R., {Griffin}, M., {Gundersen}, J., {Halpern}, M.,
  {Hargrave}, P., {Hughes}, D., {Klein}, J., {Marsden}, G., {Martin}, P.,
  {Mauskopf}, P.~D., {Netterfield}, B., {Olmi}, L., {Pascale}, E., {Rex}, M.,
  {Scott}, D., {Semisch}, C., {Truch}, M., {Tucker}, C., {Tucker}, G.,
  {Turner}, A.~D., \& {Weibe}, D. 2004, in Presented at the Society of
  Photo-Optical Instrumentation Engineers (SPIE) Conference, Vol. 5498,
  Millimeter and Submillimeter Detectors for Astronomy II. Edited by Jonas
  Zmuidzinas, Wayne S. Holland and Stafford Withington Proceedings of the SPIE,
  Volume 5498, pp. 42-54 (2004)., ed. J.~Zmuidzinas, W.~S. Holland, \&
  S.~Withington

\bibitem[{{Griffin} {et~al.}(2002){Griffin}, {Bock}, \& {Gear}}]{griffin02}
{Griffin}, M.~J., {Bock}, J.~J., \& {Gear}, W.~K. 2002, \ao, 41, 6543

\bibitem[{{Griffin} {et~al.}(2003){Griffin}, {Swinyard}, \& {Vigroux}}]{grif03}
{Griffin}, M.~J., {Swinyard}, B.~M., \& {Vigroux}, L.~G. 2003, in Presented at
  the Society of Photo-Optical Instrumentation Engineers (SPIE) Conference,
  Vol. 4850, IR Space Telescopes and Instruments. Edited by John C. Mather .
  Proceedings of the SPIE, Volume 4850, pp. 686-697 (2003)., ed. J.~C.
  {Mather}, 686--697

\bibitem[{{Hargrave} {et~al.}(2006){Hargrave}, {Waskett}, {Lim}, \&
  {Swinyard}}]{hargrave06}
{Hargrave}, P., {Waskett}, T., {Lim}, T., \& {Swinyard}, B. 2006, in Presented
  at the Society of Photo-Optical Instrumentation Engineers (SPIE) Conference,
  Vol. 6275, Millimeter and Submillimeter Detectors and Instrumentation for
  Astronomy III. Edited by Zmuidzinas, Jonas; Holland, Wayne S.; Withington,
  Stafford; Duncan, William D.. Proceedings of the SPIE, Volume 6275, pp.
  627514 (2006).

\bibitem[{{Holland} {et~al.}(2006){Holland}, {MacIntosh}, {Fairley}, {Kelly},
  {Montgomery}, {Gostick}, {Atad-Ettedgui}, {Ellis}, {Robson}, {Hollister},
  {Woodcraft}, {Ade}, {Walker}, {Irwin}, {Hilton}, {Duncan}, {Reintsema},
  {Walton}, {Parkes}, {Dunare}, {Fich}, {Kycia}, {Halpern}, {Scott}, {Gibb},
  {Molnar}, {Chapin}, {Bintley}, {Craig}, {Chylek}, {Jenness}, {Economou}, \&
  {Davis}}]{holland06}
{Holland}, W., {MacIntosh}, M., {Fairley}, A., {Kelly}, D., {Montgomery}, D.,
  {Gostick}, D., {Atad-Ettedgui}, E., {Ellis}, M., {Robson}, I., {Hollister},
  M., {Woodcraft}, A., {Ade}, P., {Walker}, I., {Irwin}, K., {Hilton}, G.,
  {Duncan}, W., {Reintsema}, C., {Walton}, A., {Parkes}, W., {Dunare}, C.,
  {Fich}, M., {Kycia}, J., {Halpern}, M., {Scott}, D., {Gibb}, A., {Molnar},
  J., {Chapin}, E., {Bintley}, D., {Craig}, S., {Chylek}, T., {Jenness}, T.,
  {Economou}, F., \& {Davis}, G. 2006, in Presented at the Society of
  Photo-Optical Instrumentation Engineers (SPIE) Conference, Vol. 6275,
  Millimeter and Submillimeter Detectors and Instrumentation for Astronomy III.
  �Edited by Zmuidzinas, Jonas; Holland, Wayne S.; Withington, Stafford;
  Duncan, William D.. �Proceedings of the SPIE, Volume 6275, pp. 62751E
  (2006).

\bibitem[{{Hughes} {et~al.}(2002){Hughes}, {Aretxaga}, {Chapin},
  {Gazta{\~n}aga}, {Dunlop}, {Devlin}, {Halpern}, {Gundersen}, {Klein},
  {Netterfield}, {Olmi}, {Scott}, \& {Tucker}}]{hughes02}
{Hughes}, D.~H., {Aretxaga}, I., {Chapin}, E.~L., {Gazta{\~n}aga}, E.,
  {Dunlop}, J.~S., {Devlin}, M.~J., {Halpern}, M., {Gundersen}, J., {Klein},
  J., {Netterfield}, C.~B., {Olmi}, L., {Scott}, D., \& {Tucker}, G. 2002,
  \mnras, 335, 871

\bibitem[{{Lamarre} {et~al.}(1998){Lamarre}, {Giard}, {Pointecouteau},
  {Bernard}, {Serra}, {Pajot}, {D{\'e}sert}, {Ristorcelli}, {Torre}, {Church},
  {Coron}, {Puget}, \& {Bock}}]{lam98}
{Lamarre}, J.~M., {Giard}, M., {Pointecouteau}, E., {Bernard}, J.~P., {Serra},
  G., {Pajot}, F., {D{\'e}sert}, F.~X., {Ristorcelli}, I., {Torre}, J.~P.,
  {Church}, S., {Coron}, N., {Puget}, J.~L., \& {Bock}, J.~J. 1998, \apjl, 507,
  L5

\bibitem[{{Lasker} {et~al.}(1987){Lasker}, {Jenkner}, \& {Russell}}]{lasker87}
{Lasker}, B.~M., {Jenkner}, H., \& {Russell}, J.~L. 1987, NASA STI/Recon
  Technical Report N, 88, 30547

\bibitem[{{Mather}(1984)}]{mather_1984}
{Mather}, J.~C. 1984, \ao, 23, 584

\bibitem[{Mortari {et~al.}(2001)Mortari, Junkins, \& Samaan}]{mort01}
Mortari, D., Junkins, J.~L., \& Samaan, M.~A. 2001, Lost--in--{S}pace {P}yramid
  {A}lgorithm for {R}obust {S}tar {P}attern {R}ecognition, AAS Paper 01--004 of
  the 22th Annual AAS Rocky Mountain Guidance and Control

\bibitem[{{Olmi}(2001)}]{olmi01}
{Olmi}, L. 2001, Int. J. of Infrared and Millim. Waves, 22, 791

\bibitem[{{Olmi}(2002)}]{olmi02}
{Olmi}, L. 2002, in Presented at the Society of Photo-Optical Instrumentation
  Engineers (SPIE) Conference, Vol. 4849, Highly Innovative Space Telescope
  Concepts Edited by Howard A. MacEwen. Proceedings of the SPIE, Volume 4849,
  pp. 245-256 2002., ed. H.~A. {MacEwen}, 245--256

\bibitem[{{Olmi}(2007)}]{olmi07}
---. 2007, Applied Optics, Accepted for publication, 791

\bibitem[{{Patanchon} {et~al.}(2008)}]{patanchon2007}
{Patanchon}, G. {et~al.} 2008, ApJ, accepted

\bibitem[{{Rex} {et~al.}(2006){Rex}, {Chapin}, {Devlin}, {Gundersen}, {Klein},
  {Pascale}, \& {Wiebe}}]{rex06}
{Rex}, M., {Chapin}, E., {Devlin}, M.~J., {Gundersen}, J., {Klein}, J.,
  {Pascale}, E., \& {Wiebe}, D. 2006, in Presented at the Society of
  Photo-Optical Instrumentation Engineers (SPIE) Conference, Vol. 6269,
  Ground-based and Airborne Instrumentation for Astronomy. Edited by McLean,
  Ian S.; Iye, Masanori. Proceedings of the SPIE, Volume 6269, pp. 62693H
  (2006).

\bibitem[{{Rownd} {et~al.}(2003){Rownd}, {Bock}, {Chattopadhyay}, {Glenn}, \&
  {Griffin}}]{rownd03}
{Rownd}, B., {Bock}, J.~J., {Chattopadhyay}, G., {Glenn}, J., \& {Griffin},
  M.~J. 2003, in Presented at the Society of Photo-Optical Instrumentation
  Engineers (SPIE) Conference, Vol. 4855, Millimeter and Submillimeter
  Detectors for Astronomy. Edited by Phillips, Thomas G.; Zmuidzinas, Jonas.
  Proceedings of the SPIE, Volume 4855, pp. 510-519 (2003)., ed. T.~G.
  {Phillips} \& J.~{Zmuidzinas}, 510--519

\bibitem[{{Truch} {et~al.}(2008)}]{truch2007}
{Truch}, M. {et~al.} 2008, ApJ, accepted

\bibitem[{Truch(2007)}]{truch-thesis}
Truch, M. D.~P. 2007, {PhD} dissertation, Brown University, Department of
  Physics

\bibitem[{{Turner} {et~al.}(2001){Turner}, {Bock}, {Beeman}, {Glenn},
  {Hargrave}, {Hristov}, {Nguyen}, {Rahman}, {Sethuraman}, \&
  {Woodcraft}}]{turner01}
{Turner}, A.~D., {Bock}, J.~J., {Beeman}, J.~W., {Glenn}, J., {Hargrave},
  P.~C., {Hristov}, V.~V., {Nguyen}, H.~T., {Rahman}, F., {Sethuraman}, S., \&
  {Woodcraft}, A.~L. 2001, \ao, 40, 4921

\end{thebibliography}

\clearpage
%%%%%%%%%%%%%%%%%%%%%%%%%%%%%%%%%%%%%%%%%%%%%%%%%%%%%
\begin{deluxetable}{lcc}
\tablewidth{0pt}
\small
\tablecaption{\blast\ optics design summary. \label{table:blast_optics}}
\tablehead{ \colhead{Element} & \colhead{\blastkir} & \colhead{\blastmcm} }
\startdata
		Primary Mirror Diameter 	& 2\,m 	& 1.8\,m \\
		Effective Focal Length   		& 10\,m	& 9\,m   \\
		Field Taper      	& --15\,db	& --7.5\,db     \\
		Obscuration     	& 14\%	& 7\%     \\
		FWHM 250\,\micron 	& 40\arcsec	& 30\arcsec	\\
		FWHM 350\,\micron 	& 58\arcsec	& 42\arcsec	\\
		FWHM 500\,\micron 	& 75\arcsec	& 60\arcsec	\\ 
\enddata
\tablecomments{The \blastkir\ instrument used a larger mirror which was under-illuminated. In order to maintain the point-source detectability, the \blastmcm\ telescope was more aggressively illuminated. The total transmission of the optical system was evaluated with optical CAD software and using the known efficiency of each smooth-walled feed horn-coupled pixel.}
\end{deluxetable}%

\begin{deluxetable}{lccc}
\tablewidth{0pt} 
\small 
\tablecaption{Summary of relevant detector characteristics. \label{table:det_params}}
\tablehead{ & \colhead{250\,\micron} & \colhead{350\,\micron}  & \colhead{500\,\micron}}
\startdata
		Light detectors	& 139 & 88 & 43 \\
		Dark pixels	&  2   & 2   &  2  \\
		Resistors       &  1   & 1   &  1  \\
		Thermistors    &  2   & 2   &  2 \\
		G (pW/K)       &  880   &  640  & 480  \\ 
		$\Delta (K)$  	& \multicolumn{3}{c}{50}    \\
		$R_0$ ($\Omega$)     	& \multicolumn{3}{c}{55}    \\ 
		$R_{\rm L}$ (M$\Omega$)     	& \multicolumn{3}{c}{$7 + 7$}   \\ 
		$\tau$ (ms)		& \multicolumn{3}{c}{$2$}         \\
		Temperature (mK)        & \multicolumn{3}{c}{270}\\
		NEP (W\,Hz$^{-1/2}$)    & \multicolumn{3}{c}{$3 \times 10^{-17}$}\\
\enddata
\tablecomments{Bolometer parameters are given for a typical array pixel. $\Delta$ and $R_{\rm o}$ define the bolometric model, $R(T) = R_0\,e^{\sqrt{\Delta/T}}$, while the optical time constant, $\tau$, and the thermal conductance, $G$,  define its thermal behavior. The NEP is computed at 1\,Hz.}
\end{deluxetable}

\begin{deluxetable}{lc}
\tablewidth{0pt} 
\small 
\tablecaption{Weights of the main components on the payload.\label{tab:mass}}
\tablehead{  &  \colhead{Weight} \\
	\colhead{Component} & \colhead{(kg)} }
\startdata
{\bf Inner Frame}  	& {\bf 705} \\ 
\hspace{.25in}Frame               	& 120  \\
\hspace{.25in}Mirror             	& 115  \\
\hspace{.25in}Cryostat (empty)    	& 215  \\
\hspace{.25in}Electronics        	& 65 \\
\hspace{.25in}Cryogens            	& 40   \\
{\bf Outer Frame}            		& {\bf 1100} \\
\hspace{.25in}Batteries               	& 80  \\
\hspace{.25in}Solar Panels             	& 30  \\
\hspace{.25in}Electronics              	& 110  \\
\hspace{.25in}Sun Shields              	& 195  \\
\hspace{.25in}Frame                 	& 195  \\
{\bf CSBF Hardware \& Electronics}    	& {\bf 215}\\[2ex]
{\bf\large Total Weight at Launch}     	& {\bf\large 2020} 
\enddata
\tablecomments{Bold face values refer to the \blastmcm\ configuration and include the items shown here as well as all the other elements on the gondola. The total weight at launch was measured at the hook of the launch vehicle before take-off, and the moment of inertia was ${\sim}$\,4500\,kg\,m$^2$.}
\end{deluxetable}

\begin{deluxetable}{lrcrcrccc}
\tablewidth{0pt} 
\small 
\tablecaption{\blast\ temperatures \label{tab:thermal} (\degree\/C).}
\tablehead{  & \multicolumn{2}{c}{\blastfs} &  \multicolumn{2}{c}{\blastkir}& \multicolumn{2}{c}{\blastmcm} & \\
& \colhead{Trop} & \colhead{Diurnal} & \colhead{Trop} & \colhead{Diurnal} & \colhead{Trop} & \colhead{Diurnal} & \colhead{Model}
}
\startdata
ACS 5V DCDC\tablenotemark{a} & 25& 51 to
23& 15& 62 to 46& 3.6& 41 to 36& ... \\
ACS Case\tablenotemark{b}& ... & ... & --30& 21 to 1& --21& 16 to 10& 28 to 13 \\
PV Case\tablenotemark{c}& ... & ... & --21& 21 to 13& --21& 17 to 12& 18 to 4 \\
PV Air& --5& 26 to 1& --7& 35 to 24& --6& 24 to 29& ... \\
Preamp Case\tablenotemark{d}& --20& 10 to --10& ... & 31 to 19& --5& 25 to 21& 25 to 19 \\
Inner Frame& --16&-2 to --27& --22& 14 to --2& --32& --2 to --5& ... \\
Batteries\tablenotemark{f}& 10&17 to 10&-3& 21 to 10& --6& 19 to 17& 16 to 0 \\
Solar Panels\tablenotemark{g}& ... & ... & --30& 66 to 5& --29& 79 to 58& 85 \\
Mylar Shields &--30& 17 to --55& --34& 18 to --4& --38& 20 to 10\tablenotemark{h}& ...\\
Primary Mirror& --24& --8 to --38& ... & 8 to --15& --17& --6 to --9& --25 to --10\\
Secondary Mirror& --33& 0 to --70& ... & ... & --26& --13 to --16\tablenotemark{i}& ... \\
Pivot&
--33& 15 to --50& --38& 20 to --5& --35& 57 to 36\tablenotemark{j}& ... \\
Outer Frame\tablenotemark{k}& --11&
--1.4 to --37& --30& 5 to --13& --24& 2 to --3& 25 to --5 \\
Star Camera& ... & ... & --11& 15 to 1& --12& 6 to 1& 18 to 5 \\
Sun Sensor\tablenotemark{l}& 0& 31& --23&
43 to 30& ... & ... & ... \\

\enddata
\tablecomments{\blast\ temperatures in the three flights, and as predicted by a Thermal Desktop (http://www.thermaldesktop.com/) model for \blastmcm\@.  The model predicted larger diurnal variations than observed, partially because \blastmcm\ followed a very southerly track, where the Sun elevation varied less than the simulation's track.  The \textit{Trop} column gives the minimum temperature the component reached during ascent through the Tropopause.  The \textit{Diurnal} column gives the daily range of temperatures.  In some cases, there was slowly increasing temperature over and above the diurnal variation due to UV degradation of the Mylar shields.}
\tablenotetext{a}{The inside of the case was painted white for \blastmcm.}
\tablenotetext{b}{Partially painted white.}
\tablenotetext{c}{Painted white.}
\tablenotetext{d}{Painted white + cooling loop.}
\tablenotetext{f}{Packaged in a foam box.}
\tablenotetext{g}{The model neglects current draw.}
\tablenotetext{h}{Increased by 20\degree\/C over 12 days.}
\tablenotetext{i}{Increased by 5\degree\/C over 12 days.}
\tablenotetext{j}{Increased by 4\degree\/C over 12 days}
\tablenotetext{k}{Measured at the lock motor near the elevation bearing.}
\tablenotetext{l}{With cooling pump to heat exchanger.}
\end{deluxetable}

\begin{deluxetable}{lc}
\tablewidth{0pt} 
\small 
\tablecaption{The steady--state power budget for \blast\@. \label{table:power}}
\tablehead{ \colhead{Component} & \colhead{Power Consumption}  \\
& \colhead{(W)}}
\startdata
\multicolumn{2}{l}{\bf ACS SIDE} \\
 \hspace{.25in}Pointing Motors\tablenotemark{a}     &  85\\
  \hspace{.25in}Flight Computers\tablenotemark{b}   &  55\\
  \hspace{.25in}Star Cameras       &  55 \\
  \hspace{.25in}ACS\tablenotemark{c}                &  30  \\
  \hspace{.25in}Gyro Heaters\tablenotemark{d}       &  30\\
  \hspace{.25in}Aux. Systems       &  15  \\
  \hspace{.25in}Differential GPS   &   5\\
  \hspace{.25in}BLAST06 Sun Sensor &   5\\
  \hspace{.25in}LOS Transmitters   & 150 \\[2ex]
ACS TOTAL             & 430\\[3ex]

\multicolumn{2}{l}{\bf DAS SIDE} \\
  \hspace{.25in}DAS                   &113\\
  \hspace{.25in}Preamps \& Cryostat    &113\\[2ex]
DAS TOTAL               &225\\
\enddata
\tablecomments{
 When slewing, the pointing
  motors may have short-term excursions up to 560 Watts, nearly doubling
  \blast's power consumption.  The LOS Transmitters are only on at the
  start of the flight and do not contribute to the steady-state budget.}
\tablenotetext{a}{Steady state}
\tablenotetext{b}{Includes everything else in the Pressure Vessel}
\tablenotetext{c}{ACS crate only}
\tablenotetext{d}{Mean value}
\end{deluxetable}

 \end{document}